\documentclass[lettersize,journal]{IEEEtran}
\usepackage{amsmath,amsfonts}
\usepackage{array}
\usepackage[caption=false,labelfont=sf,textfont=sf]{subfig}
\usepackage{textcomp}
\usepackage{stfloats}
\usepackage{url}
\usepackage{verbatim}
\usepackage{graphicx}
\usepackage{cite}
\makeatletter

\newcommand{\Rmnum}[1]{\expandafter\@slowromancap\romannumeral #1@}
\makeatother

\usepackage{CJK}
\usepackage{indentfirst}
\usepackage{cases}

\usepackage{amssymb}
\usepackage{enumerate}
\usepackage{graphicx}
\usepackage{amsmath}
\usepackage{bm}
\usepackage{amsfonts}
\usepackage{amsmath}
\usepackage{algorithmic}
\usepackage[ruled]{algorithm2e}
\usepackage{textcomp}
\usepackage{color,xcolor}
\usepackage{indentfirst}
\usepackage{amsthm}
 \usepackage{epstopdf}
\newtheorem{Lemma}{Lemma}
\newtheorem{mypro}{Proposition}
\begin{document}
%
\title{Energy Consumption Minimization in Secure Multi-antenna UAV-assisted MEC Networks with Channel Uncertainty}
\author{Weihao Mao,~Ke~Xiong,~\IEEEmembership{Member,~IEEE}, ~Yang Lu,~\IEEEmembership{Member,~IEEE},\\Pingyi Fan,~\IEEEmembership{Senior Member,~IEEE}, and Zhiguo Ding,~\IEEEmembership{Fellow,~IEEE}\\
\thanks{This work was supported in part by the Fundamental Research Funds for the Central Universities under Grant 2022JBQY004, in part by the National Natural Science Foundation of China (NSFC) under Grant 62101025, in part by China Postdoctoral Science Foundation under Grant BX2021031 and 2021M690342 and in part by Beijing Nova Program under Grant Z211100002121139. \emph{(Corresponding author: Yang Lu.)}}
\thanks{W. H. Mao,  K. Xiong, and Y. Lu, are with the Engineering Research Center of Network Management Technology for High Speed Railway of Ministry of Education, School of Computer and Information Technology, Beijing Jiaotong University, Beijing 100044, China (e-mail: 21120382\{kxiong\}yanglu@bjtu.edu.cn).}
\thanks{P. Y. Fan is with the Beijing National Research Center for Information Science and Technology, and also with the Department of Electronic Engineering, Tsinghua University, Beijing 100084, China (e-mail: fpy@tsinghua.edu.cn).}
\thanks{Z. G. Ding is with the School of Electrical and Electronic Engineering, The University of Manchester, Manchester M13 9PL, U.K. (e-mail: zhiguo.ding@manchester.ac.uk).}
}
\maketitle

\begin{abstract}
This paper investigates the robust and secure task transmission and computation scheme in multi-antenna unmanned aerial vehicle (UAV)-assisted mobile edge computing (MEC) networks, where the UAV is dual-function, i.e., aerial MEC and aerial relay. The channel uncertainty is considered during information offloading and downloading. An energy consumption minimization problem is formulated under some constraints including users' quality of service and information security requirements and the UAV's trajectory's causality, by jointly optimizing the CPU frequency, the offloading time, the beamforming vectors, the artificial noise and the trajectory of the UAV, as well as the CPU frequency, the offloading time and the transmission power of each user. To solve the non-convex problem, a reformulated problem is first derived by a series of convex reformation methods, i.e., semi-definite relaxation, S-Procedure and first-order approximation, and then, solved by a proposed successive convex approximation (SCA)-based algorithm. The convergence performance and computational complexity of  the proposed algorithm are analyzed. Numerical results demonstrate that the proposed scheme outperform existing benchmark schemes. Besides, the proposed SCA-based algorithm is superior to traditional alternative optimization-based algorithm.

\end{abstract}

%
\IEEEpeerreviewmaketitle

\setlength{\parindent}{1em}
\section{Introduction}
\noindent A. {\it{Background}}\\

With the rapid development of the Internet of Things (IoT), there has been an explosive growth in the number of smart devices [\ref{xian1}], which support many novel intelligent applications such as face recognition and augmented reality. Generally, the intelligent applications require the devices to handle lots of computation-intensive and latency-sensitive tasks [\ref{xian2}]. However, due to the limited computation capability, the devices may not be able to handle the computation tasks, which thus, degrades the quality of service (QoS) of the applications [\ref{xian3}]. As a remedy, the mobile edge computing (MEC) has been proposed as a promising technology [\ref{xian4}]. In particular, the computation tasks are offloaded to and handled at the powerful MEC nodes, which then, feedback the results to the devices. Since the MEC nodes are usually deployed around the devices, the computation tasks can be finished timely with relatively high-quality wireless coverage [\ref{xian5}]. Besides, the open nature of wireless communications brings the risks of information leakage. Therefore, guaranteeing information security is of high significance in MEC networks\cite{1}.    \\
\indent
On the other hand, the unmanned aerial vehicle (UAV) has been adopted as a wireless coverage enabler for various wireless communication networks, which is able to establish high-quality line-of-sight (LoS) link due to its flexible mobility \cite{2}. \textcolor[rgb]{0,0,0}{Strong LOS link generally leads to sparse channel and improved performance such as channel estimation. A novel method of measuring channel sparsity is firstly proposed in \cite{A1}, and the relation between channel sparsity and link performance is revealed.} Besides, to further improve the spectral efficiency, the multi-antenna UAV has drawn increasing attention \cite{3}. However, due to the flight vibration of the UAV, the channel error is inevitable in channel estimation of the link between the UAV and the ground user/node, which makes the robust design crucial in UAV-assisted communication networks \cite{4}. \textcolor[rgb]{0,0,0}{A pioneering work is conducted in \cite{A2}, which firstly reveals that mobility and vibration of UAV have significant impacts on channel non-stationarity and space-time-frequency correlation characteristics.}   \\
\indent
Recently, a trend to employ the UAV in the MEC networks has been reported, especially when some emergencies happen, such as earthquakes and flood disaster, which cause the terrestrial MEC networks to be overloaded or even unusable [\ref{xian14}]. Typically, there are two kinds of UAV-assisted MEC networks, where the UAV is respectively utilized as aerial MEC and aerial relay \cite{bg1}, \cite{bg2}. Furthermore, the UAV is designed to realize dual functions, i.e., MEC and relay, to provide more various services \cite{bg3}. However, due to the information security issue in the MEC networks and the channel uncertainty in the UAV communications, designing secure UAV-assisted MEC networks with channel uncertainty is of significance. \\

\noindent B. {\it{Related Work}}\\
\\
\indent
So far, there have been many existing works on  UAV-assisted MEC networks, see e.g., \cite{mec1, mec2, relay1, relay2, mec_relay1, mec_relay2, mec_relay3, mec_relay4}. In \cite{mec1} and \cite{mec2}, the bits of computation tasks and sum computation rates were respectively maximized for UAV-assisted networks where the UAV served as an aerial MEC. Since the UAV may be overloaded with lots of computation tasks, some works utilized the UAV as an aerial relay between the powerful ground MEC and the users. In \cite{relay1} and \cite{relay2}, the energy consumption minimization problem was studied for UAV-relay networks with a multi-antenna ground MEC and distributed MECs, respectively. To inherit the advantages of the UAV as aerial MEC and aerial relay, some works focused on the dual-function UAV. In \cite{mec_relay1}, the weighted sum of total energy consumption of UAV and the total service delay of users was minimized in a UAV-assisted MEC network where the coordinate of the UAV was fixed. In \cite{mec_relay2}, an energy consumption minimization design was proposed for a UAV-assisted MEC network, where the computation resource allocation, bandwidth scheduling and the trajectory of the UAV were optimized. In \cite{mec_relay3}, a wireless powered UAV-assisted MEC network was investigated, where the UAV forwarded both information and energy to users, and the bits of computation tasks were maximized by optimizing computation resource allocation, time slot scheduling and the trajectory of the UAV. In \cite{mec_relay4}, a multi-antenna dual-function UAV was considered and the energy consumption was minimized by optimizing the beamforming vectors, the CPU frequency and the trajectory of UAV, and transmission power and the CPU frequency of users.
\\
\indent
However, the information security was not considered in the above mentioned works, which may degrade the QoS of users in MEC networks. Thus, some works studied the secure transmission design in UAV-assisted MEC networks, see, e.g. \cite{sec1, sec2, sec3}. In \cite{sec1}, the max-min secure computation rate problem was studied for a dual-UAV assisted MEC networks with one serving UAV and one jamming UAV. In \cite{sec2}, the computation rate was maximized for a non-orthogonal multiple access (NOMA)-based UAV-assisted MEC network where a ground jammer was employed to confuse the aerial eavesdropper. In \cite{sec3}, a max-min computation rate problem was investigated for a UAV-assisted MEC network where the UAV was equipped with two antennas for receiving the offloading data and sending jamming signals, respectively. Besides the jamming scheme, the multi-antenna UAV is able to provide abundant spatial degree of freedom to achieve physical-layer security \cite{sec4}, which has been widely used in UAV-assisted networks.  \\
\indent
It is noticed that in \cite{mec1, mec2, relay1, relay2, mec_relay1, mec_relay2, mec_relay3, mec_relay4, sec1, sec2, sec3, sec4}, the perfect channel state information (CSI) assumption was adopted. Due to the mobility and vibration of the UAV, the perfect CSI assumption may be overly ideal. Thus, some works investigated the robust design for UAV communications with channel uncertainty, see, e.g., \cite{rob1, rob3, rob4}. In \cite{rob1}, the energy efficiency maximization problem was studied for a NOMA-based UAV network with the imperfect CSI between the UAV and users. In \cite{rob3}, the secrecy energy efficiency maximization problem was studied for a dual-UAV network, where the channel error was taken into account for the links associated the serving UAV. In \cite{rob4}, a robust transmission scheme was proposed for a UAV-assisted MEC network to minimize the total energy consumption with a given UAV's trajectory.
\\

\noindent C. {\it{Contributions}}\\

\indent
{\it To the authors' best knowledge, the secure multi-antenna UAV-assisted MEC network with channel uncertainty has not been studied thus far}. It is worth to note that channel uncertainty is unavoidable due to the movement and vibration of the UAV and the information security is a central consideration in the MEC networks. {\it To fill this gap}, in this paper, we investigate the robust and secure task transmission and computation scheme in multi-antenna UAV-assisted MEC networks, where the UAV is with both MEC and relay functions and the channel uncertainty is considered during offloading and downloading. The main contributions are summarized as follows: \\
\indent
1) An energy consumption minimization problem is formulated for the UAV-assisted MEC network, under constraints of the QoS and information security requirements at users and the causality of the UAV’s trajectory, by jointly optimizing the CPU frequency, the offloading time, the beamforming vectors, the artificial noise (AN) and the trajectory of the UAV, and the CPU frequency, the offloading time and the transmission power of each user. Different from previous investigations, imperfect CSI is considered in this paper, while the computation results are protected from eavesdropping. \\
\indent
2) To tackle the non-convex problem, an algorithm is proposed to optimize all variables jointly. A convex approximation problem of the considered problem is first presented via a series of convex reformulation methods, including semi-definite relaxation (SDR), S-Produce and the first-order approximation. Then, a successive convex approximation (SCA)-based algorithm is designed to improve the approximation precision. Furthermore, the convergence of the proposed algorithm is theoretically proved. Different from traditional alternative optimization (AO)-based algorithm which optimizes the coupling variables separately, the proposed algorithm jointly optimize all variables. \\
\indent
3) Numerical results validate the proposed algorithm. It is also shown that the proposed scheme outperforms some existing benchmark schemes.  By considering time slot scheduling and multi-antenna UAV, the overall energy consumption is saved. Besides, the robust design guarantees the computation tasks to be finished under channel uncertainty. Moreover, the proposed algorithm is superior to the existing AO-based algorithm in term of total energy consumption.  \\
\indent
The rest of the paper is organized as follows. Section \Rmnum{2} presents the system model and the optimization problem. Section \Rmnum{3} reformulates and solves the considered problem. Section \Rmnum{4} presents the numerical results. Finally, in Section \Rmnum{5}, we conclude the paper.\\
\indent
{\it Notations}: Throughout this paper, $\bf X$, $\bf x$, $x$ are respectively denoted as matrix, vector, scalar. ${\rm Tr}(\bf X)$ denotes the trace of ${\bf X}$. ${\rm Re}\{ x \}$ denotes the real part of $x$. $\vert \cdot \vert$ and $\Vert \cdot \Vert$  denote the absolute value of a complex scalar and the 2-norm of a complex vector, respectively. ${\bf h}^T$ and ${\bf h}^H$ denote the transpose and conjugate transpose of ${\bf h}$, respectively. $\mathbb{C}^{M}$ and  $\mathbb{C}^{M \times N}$ denote the set of $M$ complex-valued vectors and $M \times N$ complex-valued matrices, respectively. $\otimes$ denotes the Kronecker product.

\section{System Model}
We consider a UAV-assisted MEC network as shown in Fig. 1, where a UAV MEC and a ground MEC help $K$ ground users (GUs) to finish their computation tasks in presence of $M$ eavesdroppers. The UAV is equipped with $N_{\mathrm T} = N_{\rm x} \times N_{\rm y}$ uniform planar array (UPA) antennas while the GUs, the eavesdroppers and the ground MEC are equipped with single-antenna. For clarity, let $\mathcal{K} \triangleq \lbrace 1,2,..,K \rbrace $ and $\mathcal{M} \triangleq \lbrace 1,2,..,M \rbrace $ denote the set of GUs and eavesdroppers, respectively. Due to the limited computation capability and the battery capacity, the GUs offload partial computation tasks to the UAV MEC and the ground MEC with assistance of the UAV. It is noted that in our considered system, the UAV is with dual functions. Firstly, it serves as an aerial MEC to handle the computation tasks from the GUs locally. Secondly, once the UAV is overloaded, it forwards partial computation tasks to the ground MEC like an aerial relay. During the download period, the GUs receive the computation results from the UAV, while the eavesdroppers may intercept the GUs.\\

\begin{figure}[t]
\begin{center}
\centerline{\includegraphics[ width=.5\textwidth]{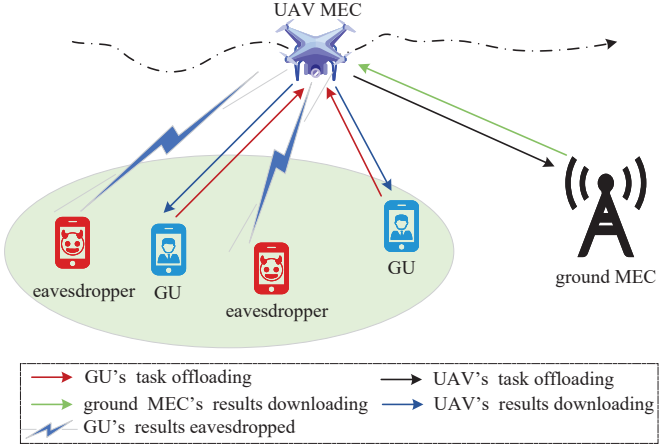}}
\caption{System model.}
\label{sys}
\end{center}
\end{figure}

\noindent A. {\it{UAV Trajectory Model and Channel Model}}\\

\noindent {\it 1) ~UAV Trajectory}:\\
\indent
The locations of GUs, eavesdroppers and the ground MEC are assumed to be fixed with vertical coordinate as $0$, and their horizontal coordinates are respectively denoted as ${\bf u}_k = {(x_{k}, y_{k})}$, ${\bf u}_{\mathrm{Eve},m} = {(x_{\mathrm{Eve},m}, y_{\mathrm{Eve},m})}$ and ${\bf u}_{\rm a} = {(x_{ \rm a}, y_{\rm a})}$. \\
\indent
Let $T$ and $H$ denote the flight period and the flight altitude of the UAV, respectively. Similar with \cite{mec_relay4}, the discrete path planning approach is adopted, where the flight period $T$ is divided into $N$ sufficiently small time slots and each time slot lasts $\Delta_{\rm T}$, i.e., $T = N\Delta_{\rm T}$.  At the $n$th time slot with $n\in{\cal N}\triangleq \{1,2,...,N\}$, the horizontal coordinate of the UAV is denoted as ${\bf q}[n]=(x_q[n],y_q[n])$. Moreover, the initial and final horizontal coordinates of the UAV are ${\bf q}_{\rm I}$ and ${\bf q}_{\rm F}$. That is,
\begin{flalign}
	{\bf q}[0] = {\bf q}_{\rm I}, {\bf q}[N+1] = {\bf q}_{\rm F}. \label{shi1}
\end{flalign}
During the $n$th time slot, the velocity of the UAV is expressed as
\begin{flalign}
	{\bf v}[n] = \frac{{\bf q}[n] - {\bf q}[n-1]}{\Delta_{\rm T}}, \forall n \in \mathcal{N}.
\end{flalign}
The corresponding propulsion power consumption [\ref{wen8}] is given by
\begin{flalign}
	P \left(  \Vert  {\bf v}[n] \Vert  \right) =~&
	P_0 \left(1 + \frac{3{\Vert  {\bf v}[n] \Vert}^2}{U_{\rm tip}^2}\right)
	+ \frac{1}{2} d_0 \rho_0 s A {\Vert  {\bf v}[n] \Vert}^3  \nonumber \\
	&+ P_{\rm H} {\left( \sqrt{1 +\frac{{\Vert  {\bf v}[n] \Vert}^4}{4 v_0^4}}-\frac{{\Vert  {\bf v}[n] \Vert}^2}{2v_0^2}
		\right)}^\frac{1}{2},
\end{flalign}
where $P_0$, $P_{\rm H}$, $U_{\rm tip}$, $\rho_0$, $A$, $d_0$ and $s$ denote the blade profile power, the induced power, the tip speed of the rotor induced velocity, the air density, the rotor disc area, the fuselage drag ratio and the rotor solidity, respectively.\\

\noindent {\it  2)~Channel Model}:\\
\indent
Taking the imperfect CSI model into account, the channel from the UAV to the $k$th GU, the $m$th eavesdropper and the ground MEC are respectively expressed as
\begin{subnumcases}
	{}
		{\bf h}_k[n] =  \bar{\bm {\mathrm h}}_k[n] + {\bf e}_k[n], \\
		\bm{\mathrm h}_{\mathrm {Eve},m}[n] = \bar{\bm{\mathrm h} }_{\mathrm{Eve},m}[n] + {\bf e}_{\mathrm {Eve},m}[n],\\
		\bm {\mathrm h}_{\rm a}[n] = \bar{\bm {\mathrm h}}_{\rm a}[n] + {\bf e}_{\rm a}[n],
\end{subnumcases}
where $\bar{\bm {\mathrm h}}_k[n] \in \mathbb{C}^{N_{\mathrm T}}$, $\bar{\bm {\mathrm h}}_{\mathrm {Eve},m}[n] \in \mathbb{C}^{N_{\mathrm T}}$  and $\bar{\bm {\mathrm h}}_{\rm a}[n] \in \mathbb{C}^{N_{\mathrm T}}$ represent the channel estimations, respectively;  ${\bf e}_k[n] \in \mathbb{C}^{N_{\mathrm T}}$, ${\bf e}_{\mathrm{Eve}, m}[n] \in \mathbb{C}^{N_{\mathrm T}}$ and ${\bf e}_{\rm a}[n] \in \mathbb{C}^{N_{\mathrm T}}$ represent the channel errors. The bounded channel error model \cite{xian10} is adopted, which means that
\begin{subnumcases}
	{}
 {\bf e}_k[n] \in \mathcal{E}_k[n] \triangleq \{ {\bf e}| {\bf e}^H{\bf C}_k[n] {\bf e} \leq 1  \},  \nonumber \\
 {\bf e}_{\mathrm{Eve},m}[n] \in \mathcal{E}_{\mathrm {Eve},m}[n] \triangleq \{ {\bf e}| {\bf e}^H{\bf C}_{\mathrm {Eve},m}[n] {\bf e}\leq 1  \}, \nonumber \\
 {\bf e}_{\rm a}[n] \in \mathcal{E}_{\rm a}[n] \triangleq \{ {\bf e} | {\bf e}^H{\bf C}_{\rm a}[n] {\bf e}\leq 1  \}  \nonumber , 
 \end{subnumcases}
 where ${\bf C}_k[n]$ $\in \mathbb{H}^{N_{\mathrm{T}}}_{+}$, ${\bf C}_{\mathrm {Eve},m}[n]$ $\in \mathbb{H}^{N_{\mathrm{T}}}_{+}$ and ${\bf C}_{\rm a}[n] $ $\in \mathbb{H}^{N_{\mathrm{T}}}_{+}$ determine the size and the shape of the error ellipsoid.\\
\indent As the channels from the UAV to the GUs, the eavesdroppers and the ground MEC are all dominated by LoS [\ref{wen3}], [\ref{wen4}]. $\bar{\bm {\mathrm h}}_k[n]$, $\bar{\bm {\mathrm h}}_{\mathrm{Eve},m}[n]$ and $\bar{\bm {\mathrm h}}_{\rm a}[n]$ are respectively expressed as
\begin{subnumcases}
	{}
		\bar{\bm {\mathrm h}}_k[n] = \frac{\sqrt{\rho} \bm {\mathrm a}_k[n]}{\sqrt{{\Vert {\bf q}[n]-{\bf u}_k \Vert}^2 + H^2}},  \\
		\bar{\bm {\mathrm h}}_{\mathrm{Eve},m}[n] = \frac{\sqrt{\rho} \bm {\mathrm a}_{\mathrm{Eve},m}[n]}{\sqrt{{\Vert {\bf q}[n]-{\bf u}_{\mathrm{Eve},m} \Vert}^2 + H^2}}, \\
		 \bar{\bm {\mathrm h}}_{\rm a}[n] = \frac{\sqrt{\rho} \bm {\mathrm a}_{\rm a}[n]}{\sqrt{{\Vert {\bf q}[n]-{\bf u}_{\rm a} \Vert}^2 + H^2}},
\end{subnumcases}
where $\rho = (\lambda_c/4\pi )^2$ with $\lambda_c$ being the wavelength of carrier frequency; $\bm {\mathrm a}_k[n] \in \mathbb{C}^{N_{\mathrm T}}$, $\bm {\mathrm a}_{\mathrm{Eve},m}[n] \in \mathbb{C}^{N_{\mathrm T}}$ and $\bm {\mathrm a}_{\rm a}[n] \in \mathbb{C}^{N_{\mathrm T}}$ are the channel vectors in the $n$th time slot from the UAV to the $k$th GU,  the $m$th eavesdropper and the ground MEC, respectively, which are given by [\ref{wen5}]
\begin{subnumcases}
	{}
	\begin{split}		
			\bm{{\mathrm a}}_k^T[n] =
			& ( 1, ..., e^{-j\frac{2\pi b f_c}{c} \sin{\varpi_k[n]} (n_{\rm x}-1) \cos{\phi_k[n]}},..., \\
			& e^{-j\frac{2\pi b f_c}{c} \sin{\varpi_k[n]} (N_{\rm x}-1) \cos{\phi_k[n]}} )
			\otimes  \\
			& ( 1, ..., e^{-j\frac{2\pi b f_c}{c} \sin{\varpi_k[n]} (n_{\rm y}-1) \sin{\phi_k[n]}}, \\
			& ...,e^{-j\frac{2\pi b f_c}{c} \sin{\varpi_k[n]} (N_{\rm y}-1) \sin{\phi_k[n]}} );
		\end{split} \\
	\begin{split}		\label{shi 1a}
		\bm{{\mathrm a}}_{\mathrm {Eve},m}^T[n] =
		& ( 1, ..., e^{-j\frac{2\pi b f_c}{c} \sin{\varpi_m^{\mathrm {Eve}}[n]} (n_{\rm x}-1) \cos{\phi_m^\mathrm {Eve}[n]}}, \\
		& e^{-j\frac{2\pi b f_c}{c} \sin{\varpi_m^{\mathrm {Eve}}[n]} (N_{\rm x}-1) \cos{\phi_m^{\mathrm {Eve}}[n]}} )
		\otimes  \\
		& ( 1, ..., e^{-j\frac{2\pi b f_c}{c} \sin{\varpi_m^\mathrm {Eve}[n]} (n_{\rm y}-1) \sin{\phi_m^\mathrm {Eve}[n]}}, \\
		& ...,e^{-j\frac{2\pi b f_c}{c} \sin{\varpi_m^\mathrm {Eve}[n]} (N_{\rm y}-1) \sin{\phi_m^\mathrm {Eve}[n]}} );
	\end{split} \\
	   \begin{split}		
	   	 \bm{{\mathrm a}}_{\rm a}^T[n] =
	   	   & ( 1, ..., e^{-j\frac{2\pi b f_c}{c} \sin{\varpi_{\rm a}[n]} (n_{\rm x}-1) \cos{\phi_{\rm a}[n]}},..., \\
	   	    & e^{-j\frac{2\pi b f_c}{c} \sin{\varpi_{\rm a}[n]} (N_{\rm x}-1) \cos{\phi_{\rm a}[n]}} )
	     	\otimes  \\
	     	& ( 1, ..., e^{-j\frac{2\pi b f_c}{c} \sin{\varpi_{\rm a}[n]} (n_{\rm y}-1) \sin{\phi_{\rm a}[n]}}, \\
	     	& ...,e^{-j\frac{2\pi b f_c}{c} \sin{\varpi_{\rm a}[n]} (N_{\rm y}-1) \sin{\phi_{\rm a}[n]}} );
	   \end{split} \label{shi 1c}
\end{subnumcases}
where $b$, $c$ and $f_c$ respectively denote the distance between the antennas, the speed of light, and the center frequency of carrier frequency. $n_{\rm x}$ and $n_{\rm y}$ are the index of the UPA's row and column, respectively. $\varpi_k[n]$, $\varpi_m^{\mathrm Eve}[n]$ and $\varpi_{\rm a}[n]$ denote the vertical angle of departure (AoD), respectively. $\phi_k[n]$, $\phi_m^{\mathrm Eve}[n]$ and $\phi_{\rm a}[n]$ denote the horizontal AoD , respectively. They are respectively denoted as
\begin{subnumcases}
	{}
		\varpi_k[n] = \arcsin{\frac{H}{\sqrt{H^2+{\Vert {\bf q}[n] - {\bf u}_k \Vert}^2}}},  \label{4a}\\
		\varpi_m^{\mathrm {Eve}}[n] = \arcsin{\frac{H}{\sqrt{H^2+{\Vert {\bf q}[n] - {\bf u}_{\mathrm{Eve},m} \Vert}^2}}}, \\
		\varpi_{\rm a}[n] = \arcsin{\frac{H}{\sqrt{H^2+{\Vert {\bf q}[n] - {\bf u}_{\rm a} \Vert}^2}}}, \\
		\phi_k[n] = \arccos{\frac{y_q[n] - y_{k}}{\Vert {\bf q}[n] - {\bf u}_k \Vert}},\\
		\phi_m^{\mathrm {Eve}}[n] = \arccos{\frac{y_q[n] - y_{\mathrm{Eve},m}}{\Vert {\bf q}[n] - {\bf u}_{\mathrm{Eve},m} \Vert}},\\
		\phi_{\rm a}[n] = \arccos{\frac{y_q[n] - y_{\rm a}}{\Vert {\bf q}[n] - {\bf u}_{\rm a} \Vert}}.\label{4d}
\end{subnumcases}
In our system, each time slot is sufficiently small, so the change of UAV's position is tiny enough, similar with [\ref{wen9}], we assume AoDs remain roughly unchanged in one time slot, so one can approximate the AoDs at the $n$th time slot with the AoDs at the end of the $(n-1)$th time slot.\\

\begin{figure}[t]
	\begin{center}
		\centerline{\includegraphics[ width=.5\textwidth]{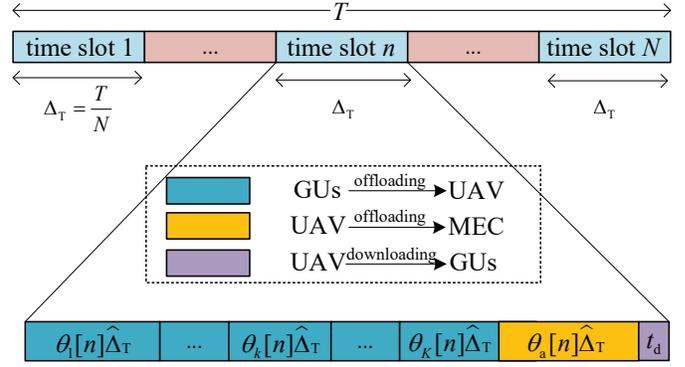}}
		\caption{Illustration of time slot division and information exchange.}
		\label{time div}
	\end{center}
\end{figure}

\noindent B. {\it{Computation and Transmission Model}} \\

\indent
To avoid the interference among GUs, similar with \cite{xian11}, each time slot period $\Delta_{\rm T}$ is divided into $(K+2)$ durations. The last duration, i.e., the $(K+2)$th duration $t_{\rm d}$, is a constant which is allocated to the UAV to gather the computation results and download them to GUs. Denote ${\widehat \Delta }_{\rm T}=\Delta_{\rm T}-t_{\rm d}$. The former $K$ durations $\{\theta_k[n] {\widehat \Delta }_{\rm T}\}$ are allocated to GUs to offload the computation tasks to the UAV in a round-robin manner and the $(K+1)$th duration $\theta_{\rm a}[n] {\widehat \Delta }_{\rm T}$ is allocated to the UAV to offload partial computation tasks to the ground MEC, where $\{\theta_k[n]\}$ and $\theta_{\rm a}[n]$ denote the utilization radio of ${\widehat \Delta }_{\rm T}$, which satisfies the following inequality
\begin{flalign}
	\sum_{k \in \mathcal{K}}{\theta_k[n]} + \theta_{\rm a}[n] \leq 1. \label{shi8}
\end{flalign}
\\

\noindent \it{1)~ GUs' Local Computing} \\
\rm \indent
At the $n$th time slot, the computation resource of the $k$th GU to handle its computation task locally is denoted as $f_{\rm l, k}[n]$. Then, the  bits of the computation task finished at the $k$th GU is
\begin{flalign}
	l_{{\mathrm l},k}[n] = \frac{f_{{\mathrm l},k}[n]}{D_k} \widehat{\Delta}_{\mathrm T},
\end{flalign}
where $D_k$ denotes the required number of CPU cycles to process one bit of the $k$th GU's computation task. According to [\ref{wen6}], the energy consumption due to local computing at the $k$th GU is expressed as
\begin{flalign}
	E_{\mathrm{l},k}[n] = v_{\mathrm l} f_{\mathrm{l},k}^3[n] \widehat{\Delta}_{\mathrm T},
\end{flalign}
where $v_{\rm l}$ denotes the effective capacitance coefficient of the processor’s chip at the $k$th GU, which is determined by the chip's architecture.\\

\noindent {\it{ 2)~GUs' Computation Task Offloading}} \\
\indent
At the $n$th time slot, the $k$th GU offloads its partial computation task to the UAV during a period of $\theta_k[n]{\widehat \Delta}_{\rm T}$. Denote the transmit power as $p_k[n]$. Then, the bits of the computation task offloaded to the UAV is
\begin{flalign}
	l_{\mathrm {o},k}[n] = \theta_k[n] \widehat{\Delta}_{\rm T} B \log_2 \left( {1+\frac{p_k[n] {\Vert \bm {\mathrm h}_k[n] \Vert}^2}{\sigma_k^2[n]}} \right),
\end{flalign}
where $\sigma_k^2[n]$ denotes the power of additive white Gaussian noise (AWGN) at the UAV , $B$ denotes the bandwidth. The energy consumption due to the computation task offloading is expressed as
\begin{flalign}
	E_{ {\mathrm o},k}[n] = \theta_k[n] \widehat{\Delta}_{\rm T} p_k[n].
\end{flalign}
\indent
Denote $I_k$ as the bits of the computation task of the $k$th GU during the UAV's flight period $T$. Similar with [\ref{xian22}], at each time slot, the computation task required to finish is denoted as $Q_k=I_k/N$. Because the imperfect CSI model is adopted, the computation task should be guaranteed to finish for the worst case. That is
\begin{flalign}
	l_{{\mathrm o},k}[n] 	\geq Q_k - l_{{\mathrm l},k}[n],\forall {\bf e}_k \in \mathcal{E}_k[n], k \in \mathcal{K}. \label{10}
\end{flalign}
Note that (\ref{10}) guarantees that the computation task of the $k$th GU required to be handled at the $n$th time slot is either finished locally  or offloaded to the UAV for all channel error cases.\\

\noindent {\it 3)~ {UAV MEC's Computing}} \\
\indent
At the $n$th time slot, the UAV MEC handles the computation tasks of all GUs during a period of $\theta_{\rm a}[n] {\widehat \Delta}_{\rm T}$, i.e., the $(K+1)$th duration. Denote the computation resource of the UAV MEC allocated to the $k$th GU as $f_{{\rm u},k}$. Then, the bits of the $k$th GU's computation task finished at the UAV is
\begin{flalign}
	l_{\mathrm u,k}[n] = \frac{f_{\mathrm u,k}[n]}{D_k} \theta_{\rm a}[n] \widehat{\Delta}_{\rm T}.
\end{flalign}
The corresponding energy consumption is expressed as [\ref{wen6}]
\begin{flalign}
	E_{\mathrm u,k}[n]  = v_{\rm u} f_{\mathrm u,k}^3[n] \theta_{\rm a}[n] \widehat{\Delta}_{\rm T},
\end{flalign}
where $v_{\rm u}$ denotes the effective capacitance coefficient of the processor's chip at the UAV MEC.\\

\noindent  {\it 4)~{UAV's Offloading}} \\
\indent
At the $n$th time slot, during the period $\theta_{\rm a}[n] \widehat{\Delta}_{\rm T}$, once the UAV MEC is overloaded, it offloads partial computation tasks to the ground MEC. Denote the transmit beamforming vector from the UAV to the ground MEC as ${\bf w}_{\rm a}\in{\mathbb C}^{N_{\rm T}}$. The bits of the computation tasks offloaded to the ground MEC is
\begin{flalign}
	l_{\rm o,\rm u}[n] = \theta_{\rm a}[n] \widehat{\Delta}_{\rm T} B \log_2 \left(
	1 + \frac{{\vert \bm {\mathrm h}_{\rm a}^H[n] {\bf w}_{\rm a}[n]\vert}^2}{\sigma_{\rm a}^2[n]}
	\right),
\end{flalign}
where $\sigma_{\rm a}^2[n]$ denotes the power of AWGN at the ground MEC. The energy consumption due to UAV's offloading is expressed as
\begin{flalign}
		E_{\rm o,\rm u} = \theta_{\rm a}[n] \widehat{\Delta}_{\rm T} {\bf w}_{\rm a}^H[n] {\bf w}_{\rm a}[n].
\end{flalign}
As the channel error ${\bf e}_{\rm a}[n]$ is involved, ${\bf w}_{\rm a}$ is required to be optimized to satisfy the following inequality, i.e.,
\begin{flalign}
	l_{\rm o,\rm u}[n] \geq \sum_{k \in \mathcal{K}}{\left( Q_k-l_{{\rm l},k}[n] - l_{{\rm u}, k}[n] \right)},~\forall  {\bf e}_{\rm a} \in \mathcal{E}_{\rm a}[n], \label{15}
\end{flalign}
where 
\begin{flalign}
	Q_k \geq l_{{\rm l}, k}[n] + l_{{\rm u},k}[n] \label{shi19},\forall k\in \mathcal{K}.
\end{flalign}
Note that (\ref{15}) guarantees that the computation tasks of all GUs offloaded to the UAV at the $n$th time slot are either finished at the UAV MEC or offloaded to the ground MEC for all channel error cases. Suppose the ground MEC is equipped with powerful computation and communication capability. The time to handle the computation tasks at the ground MEC and feedback the computation results from the ground MEC to the UAV is neglected, like [\ref{wen6}].\\

\noindent {\it{5)~Computation Results Downloading}}\\
\indent
At the $n$th time slot, the UAV sends the computation results to GUs in a multi-cast manner during a period of $t_{\rm d}$ [\ref{xian22}], i.e., the $(K+2)$th duration. During the open nature of the wireless communications, both the GUs and the eavesdroppers can receive the download signals from the UAV. The received signals at the $k$th GU and the $m$th eavesdropper are respectively expressed as
\begin{subnumcases}
	{}
		y_{k}^{\mathrm {down}}[n] =  ~\bm{\mathrm h}_k^H[n] {\bf w}_k[n] s_k[n] + n_k[n] \nonumber \\
	~~~~~~~~~~	~	  ~~       +\bm{\mathrm h}_k^H[n] \left(
			             \sum_{j \in \mathcal{K} \setminus \lbrace k \rbrace}			
			             {\bf w}_j[n] s_j[n] + {\bf z}[n]
			             \right),	\\		
	y_{\mathrm {Eve},m}^{\mathrm {down}}[n] =~ \bm {\mathrm h}_{\mathrm {Eve},m}^H[n]\left(  \sum_{j \in \mathcal{K}} {\bf w}_j[n] s_j[n] + {\bf z}[n]   \right) \nonumber  \\
	~~~~~~~~~~	~~	  ~~  +n_{\mathrm {Eve},m}[n],
\end{subnumcases}
\noindent
where ${\bf w}_k[n] \in \mathbb{C}^{N_{\rm T} }$ denotes the beamforming vector; $s_{k}[n] \in C$ represents the symbol sent by the UAV to the $k$th GU, which satisfies $\mathbb{E}\lbrace {|s_k|}^2 \rbrace=1$; ${\bf z}[n] \in\mathbb{C}^{N_{\rm T} } $ denotes the AN to confuse the eavesdroppers, which satisfies  $\mathbb{E}\lbrace {\bf z}[n]\rbrace  ={\bf 0}$ and $\mathbb{E}\lbrace {\bf z}[n] {\bf z}^H[n] \rbrace= {\bf Z}[n]$; $n_{k}[n]$ denotes AWGN at the $k$th GU with power of $\sigma_{k}^2[n]$ and $n_{\mathrm{Eve},m}[n]$ denotes AWGN at the $m$th eavesdropper with power of $\sigma_{\mathrm{Eve}, m}^2[n]$. Then the SINR at the $k$th GU and the $m$th eavesdropper for intercepting the $k$th GU is respectively given by (\ref{17a}) and (\ref{17b}).
\begin{figure*}
\begin{subnumcases}
	{}
		 \Gamma_k[n] = \frac{{\vert \bm {\mathrm h}_k^H[n] {\bf w}_k[n] \vert}^2}
		{ \sigma_{k}^2[n] + \sum\limits_{j \in \mathcal{K} \setminus \lbrace k \rbrace}
			{{\vert \bm {\mathrm h}_k^H[n] {\bf w}_j[n]\vert}^2}
			+ \bm {\mathrm h}_k^H[n] {\bf Z}[n] \bm {\mathrm h}_k[n]} , \label{17a} \\
			\Gamma_{m,k}[n] =  \frac{{\vert \bm {\mathrm h}_{\mathrm{Eve},m}^H[n] {\bf w}_k[n] \vert}^2}
		{ \sigma_{\mathrm{Eve},m}^2[n] + \sum\limits_{j \in \mathcal{K} \setminus \lbrace k \rbrace}
			{{\vert \bm {\mathrm h}_{\mathrm{Eve},m}^H[n] {\bf w}_j[n]\vert}^2}
			+ \bm {\mathrm h}_{\mathrm{Eve},m}^H[n] {\bf Z}[n] \bm {\mathrm h}_{\mathrm{Eve},m}[n]} . \label{17b}
\end{subnumcases}
\hrule
\end{figure*}
The corresponding energy consumption due to the computation results downloading is expressed as
\begin{flalign}
	E_{\rm d}[n] = t_{\rm d} \left(\sum_{k \in \mathcal{K}}{{\bf w}_k^H[n] {\bf w}_k[n]}  +{\rm Tr}\left({\bf Z}[n]\right) \right).
\end{flalign}\\

\noindent  C. {\it{Optimization Problem}}\\

\indent
The energy consumption of each GU at the $n$th time slot comprises two parts, i.e., $E_{{\rm l},k}[n]$ for local computing and $E_{{\rm o},k}[n]$ for offloading. Then, the energy consumption of all GUs is expressed as
\begin{flalign}
	E_{\rm I}[n] = \sum_{k \in \mathcal{K}}{\left(E_{\mathrm l,k}[n]+E_{\mathrm o,k}[n]\right)}.
\end{flalign}
The energy consumption of the UAV at the $n$th time slot comprises four parts, i.e., $\sum_{k \in \mathcal{K}}{E_{{\mathrm u},k}[n]}$ for computing, $E_{\rm o,\rm u}[n]$ for offloading, $ E_{\rm d}$ for sending results and $P(\Vert  {\bf v}[n] \Vert) \Delta_{\rm T}$ for flight, which is expressed as
\begin{flalign}
	E_{\rm U}[n] = \sum_{k \in \mathcal{K}}{E_{{\mathrm u},k}[n]}+E_{\rm o,\rm u}[n]
	+ E_{\rm d}[n] + P(\Vert  {\bf v}[n] \Vert) \Delta_{\rm T}.
\end{flalign}
\indent \rm
Our goal is to minimize the overall energy consumption of each time slot under constraints of the QoS and information security requirements at GUs and the causality of the UAV's trajectory. Then, the considered optimization problem is formulated as
\begin{subequations}
	\begin{align}
		\textbf{P$_1$}:&~\mathop{\min}\limits_{{\bf F} ,{\bf \Theta}, {\bf P}, {\bf W}, {\bf q}[n]}
		E_{\rm I}[n] + \eta E_{\rm U}[n]  \label{23a}
		\\
		{\rm s.t.}&~ 0 \leq f_{{\rm l}, k}[n] \leq f_{\rm l, max} \label{23b} \\
		&~ \sum\nolimits_{k \in \mathcal{K}}{f_{{\rm u},k}[n]} \leq f_{\rm u,\mathrm{max}},  \label{23c} \\
		&~ 0 \leq p_{k}[n] \leq p_{{\rm l},\mathrm{max}} ,\label{23d}\\
		&~ {\bf w}_{\rm a}^H[n] {\bf w}_{\rm a}[n] \leq p_{\rm u,\mathrm{max}},\label{23e}\\
		&~ \sum\nolimits_{k \in \mathcal{K}} {\bf w}_k^H[n] {\bf w}_k[n] + {\rm{Tr}}({\bf Z}[n])\leq p_{\rm u,\mathrm{max}},\label{23f}\\
		&~ 	\Gamma_k[n] \geq \Gamma_{\mathrm{req}},\label{23g}	\\
		&~ 	\Gamma_{m,k}[n] \leq \Gamma_{\mathrm{seq}},  \label{23h}\\
		&~ 	\Vert {\bf q}[n] - {\bf q}[n-1]\Vert \leq V_{\mathrm{max}} \Delta_{\rm T}, \label{23i}\\
		&~  \Vert {\bf q}[n] - {\bf q}_F\Vert \leq  (N-n+1)V_{\mathrm{max}} \Delta_{\rm T}, \label{23j} \\ 
		&~ \theta_k[n]\geq 0 ,\theta_{\rm a}[n]\geq 0, f_{\rm u,k}[n] \geq 0\label{23k} \\
		&~ 	{\bf Z}[n] \succeq \bm 0, \label{23l} \\
		&~ (\ref{shi1}),(\ref{shi8}), (\ref{10}), (\ref{15}),(\ref{shi19}),  \nonumber \\
		&~ \forall k \in  \mathcal{K},  m \in \mathcal{M}, \nonumber \\
		&~\forall {\bf e}_k \in \mathcal{E}_k[n], {\bf e}_{\mathrm{Eve},m} \in \mathcal{E}_{\mathrm{Eve},m}[n],{\bf e}_{\rm a} \in \mathcal{E}_{\rm a}[n], \nonumber
	\end{align}
\end{subequations}
where $\eta$ is a weighted coefficient\footnote{According to \cite{z1}, $\eta$ is employed to balance the energy consumption of the UAV and GUs, because the energy consumption of the UAV, i.e., $E_{\rm U}$, is much greater than that of GUs, i.e., $E_{\rm I}$.} 
, ${\bf F} \triangleq \{ f_{{\rm l},k}[n],f_{{\rm u},k}[n] \}$, $ {\bf \Theta} \triangleq \{ \theta_{k}[n], \theta_{\rm a}[n] \}$, ${\bf P} \triangleq \{ p_k[n] \}$ and ${\bf W} \triangleq \{ {\bf w}_k[n] , {\bf w}_{\rm a}[n],{\bf Z}[n] \}$. \\
\indent
 In (\ref{23b}) and (\ref{23c}), $f_{{\rm l},\max}$ and $f_{{\rm u},\max}$ denote the maximum
 CPU frequency of each GU and the UAV, respectively. In (\ref{23d}), (\ref{23e}) and (\ref{23f}), $p_{{\rm l},\max}$ and $p_{{\rm u},\max}$ denote the maximum transmit power of each GU and the UAV, respectively. In (\ref{23g}), $\Gamma_{\rm req}$ denotes the minimum SINR for receiving the computation results. In (\ref{23h}), $\Gamma_{\rm seq}$ denotes the secure SINR threshold to suppress the computation results leakage to eavesdroppers below a tolerable level. In (\ref{23i}), $V_{\max}$ denotes maximum velocity of the UAV and (24o) guarantees the causality of the UAV's trajectory. Besides, (\ref{23j}) guarantees that the UAV is able to reach the final point, i.e., ${\bf q}_{\rm F}$ at any time slot. Note that Problem ${\bf P}_1$ only focuses on the optimization in the $n$th time slot. Due to the association with the $(n-1)$th time slot by (25i) and the $(N+1)$th time slot by (25j), the trajectory of the UAV is derived by solving Problem ${\bf P}_1$ from $n=1$ to $n=N$.
\\
\section{Proposed Solution Approach}
\begin{figure*}[t]
	\begin{center}
		\centerline{\includegraphics[ width=\textwidth]{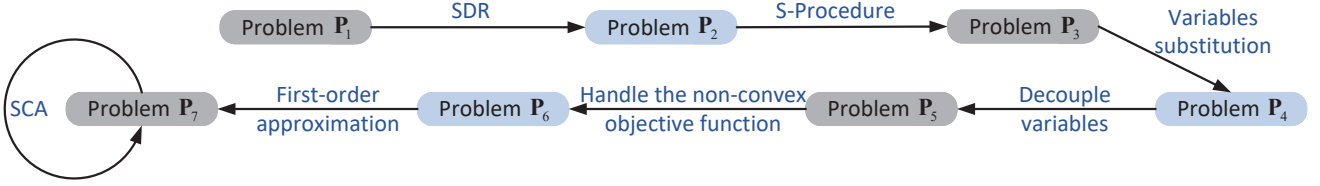}}
		\caption{The solving idea of Problem ${\bf P}_1$}
		\label{fig:idea}
	\end{center}
\end{figure*}

\indent
Thanks to the non-convex objective function (\ref{23a}) and the constraints (\ref{10}), (\ref{15}), (\ref{shi19}), (\ref{23g}) and (\ref{23h}), Problem \textbf{P$_1$} is non-convex and hard to solve. Moreover, the infinite number of channel errors makes the considered problem even computationally intractable. To tackle Problem ${\bf P}_1$, we first derive an SDR form of the considered problem, and then, employ S-Procedure to deal with channel errors. Next, a series of auxiliary variables are introduced to decouple the coupling variables and transform the non-convex objection function and constraints into convex ones via first-order approximation. At last, the SCA method is utilized to improve the approximation precision. The solving idea is summarized in Fig. \ref{fig:idea} and the detailed processes are described as follows, where for convenience we omit the time slot index $n$.\\

Define auxiliary variables $\alpha_k$ and $\alpha_{\rm a}$ satisfying that,
\begin{subequations}
	\begin{align}
		\alpha_k &\leq  p_k{\Vert \bm {\mathrm h}_k  \Vert}^2\nonumber
		\\&= p_k\left(\frac{\rho  N_{\mathrm T}}{d_k^2} + 2{\rm Re} \left\{ {\bf e}_k^H \frac{\sqrt{\rho} \bm {\mathrm a}_k}{d_k} \right\} + {\bf e}_k^H {\bf e}_k \right),
		\label{26c} \\
		{\rm and} \nonumber \\  
		\alpha_{\rm a}& \leq {\vert \bm {\mathrm h}_{\rm a}^H {\bf w}_{\rm a}\vert}^2 =
		\frac{\rho \bm {\mathrm a}_{\rm a}^H {\bf w}_{\rm a} {\bf w}_{\rm a} ^H \bm {\mathrm a}_{\rm a}}{d_{\rm a}^2} + 2{\rm Re}\left\{{\bf e}_{\rm a}^H \frac{\sqrt{\rho} {\bf w}_{\rm a} {\bf w}_{\rm a}^H  \bm {\mathrm a}_{\rm a}}{d_{\rm a}}  \right\} \nonumber \\
		&~~~~~~~~~~~~~~~~~~+ {\bf e}_{\rm a}^H {\bf w}_{\rm a} {\bf w}_{\rm a} ^H {\bf e}_{\rm a},  \label{24b}
	\end{align}
\end{subequations}
where $d_k$ and $d_{\rm a}$ are respectively defined as 
\begin{subnumcases}
	{}
	d_k \triangleq	\sqrt{\Vert {\bf q} - {\bf u}_k \Vert^2+H^2}, \\
    d_{\rm a} \triangleq	\sqrt{\Vert {\bf q} - {\bf u}_{\rm a} \Vert^2+H^2}, 
\end{subnumcases}
and then, constraints (\ref{10}) and  (\ref{15}) are respectively expressed as
\begin{subnumcases}
	{}
		\theta_k \widehat{\Delta}_{\rm T} B \log_2 \left( {1+\frac{\alpha_k}{\sigma_k^2}} \right)  \geq Q_k - l_{{\rm l},k}, \label{25a} \\
		\theta_{\rm a} \widehat{\Delta}_{\rm T} B \log_2 \left( {1+\frac{ \alpha_{\rm a}}{\sigma_{\rm a}^2}} \right)  \geq \sum\nolimits_{k \in \mathcal{K}}{(Q_k-l_{{\rm l},k} - l_{{\rm u}, k})}. \label{25b}
\end{subnumcases}
By defining that ${\bf W}_{\rm a} = {\bf w}_{\rm a} {\bf w}_{\rm a}^H$ and ${\bf W}_k = {\bf w}_k {\bf w}_k^H$, constraints (\ref{23e}), (\ref{23f}),  (\ref{23g}), (\ref{23h}) and (\ref{24b}) are respectively re-expressed as follows
\begin{subnumcases}
	{}
		\mathrm{Tr}\left( {\bf W}_{\rm a}\right)	\leq p_{\rm u,\mathrm{max}}, \label{26g} \\
		\sum\nolimits_{k \in \mathcal{K}}{{\rm Tr}({\bf W}_k)}  +{\rm Tr}({\bf Z}) \leq p_{\rm u, \mathrm{max}}, \label{26h}\\
		\sigma_{k}^2 d_k^2 + \rho \bm {\mathrm a}_k^H {\bf X}_{k,\mathrm{req}} \bm {\mathrm a}_k + 2{\mathrm{Re} \{ d_k {\bf e}_k^H \sqrt{\rho} {\bf X}_{k,\mathrm{req}}  \bm {\mathrm a}_k \}}  \nonumber \\
		+ d_k^2{\bf e}_k^H {\bf X}_{k, \mathrm{req}} {\bf e}_k  \leq 0 , \label{26i}\\
		\sigma_{\mathrm{Eve},m}^2 d_{\mathrm{Eve},m}^2  + \rho \bm {\mathrm a}_{\mathrm{Eve},m}^H {\bf X}_{k, \mathrm{seq}} \bm {\mathrm a}_{\mathrm{Eve},m} \nonumber \\
		+2{\mathrm{Re} \{ d_{\mathrm{Eve},m} {\bf e}_{\mathrm{Eve},m}^H \sqrt{\rho} {\bf X}_{k,\mathrm{seq}} \bm {\mathrm a}_{\mathrm{Eve},m} \}} \nonumber \\
		+d_{\mathrm{Eve},m}^2 {\bf e}_{\mathrm{Eve},m}^H {\bf X}_{k,\mathrm{seq}} {\bf e}_{\mathrm{Eve},m}  \geq 0, \label{26j} \\
		\alpha_{\rm a} \leq \frac{\rho \bm {\mathrm a}_{\rm a}^H {\bf W}_{\rm a} \bm {\mathrm a}_{\rm a}}{d_{\rm a}^2} + 2{\rm Re}\left\{{\bf e}_{\rm a}^H \frac{\sqrt{\rho} {\bf W}_{\rm a} \bm {\mathrm a}_{\rm a}}{d_{\rm a}}  \right\} \nonumber \\
		~~~~~~ + {\bf e}_{\rm a}^H {\bf W}_{\rm a} {\bf e}_{\rm a}, \label{26d}			
\end{subnumcases}
where $d_{{\rm Eve},m}$, ${\bf X}_{k,{\rm req}}$ and ${\bf X}_{k, {\rm seq}}$ are respectively defined as 
\begin{subnumcases}
	{}
	d_{{\rm Eve}, m} \triangleq \sqrt{\Vert {\bf q} - u_{{\rm Eve}, m} \Vert^2 +H^2 } \\
	{\bf X}_{k, {\rm req}} \triangleq {\bf Z} + \sum\nolimits_{j \in \mathcal{K} \setminus \lbrace k \rbrace} {\bf W}_j - \frac{{\bf W}_k}{\Gamma_{\rm req}}\\
	{\bf X}_{k, {\rm{seq}}} \triangleq {\bf Z} + \sum\nolimits_{j \in \mathcal{K} \setminus \lbrace k \rbrace} {\bf W}_j - \frac{{\bf W}_k}{\Gamma_{\rm seq}}
\end{subnumcases}
By dropping the rank-one constraints on on ${\bf W}_{\rm a}$ and  ${\bf W}_k$, the SDR form of Problem \textbf{P$_1$} is given by
\begin{subequations}
	\begin{align}
		\textbf{P$_2$}:&\mathop{\min}\limits_{{\bf L}_1}
		~~\tau_1\left( {\bf L}_1 \right)  \label{26a}\\
		{\rm s.t.}&~{\bf W}_{\rm a} \succeq {\bf 0}, \label{26e}\\
		&~{\bf W}_k \succeq \bf 0, \label{26f} \\	
		&~ {\rm (\ref{shi1}),(\ref{shi8}),(\ref{shi19}),(\ref{23b}),(\ref{23c}),(\ref{23d}),(\ref{23i}),(\ref{23j}),(\ref{23k}),} \nonumber \\
		&~{\rm (\ref{23l}),(\ref{26c}), (\ref{25a}),(\ref{25b}),(\ref{26g}),(\ref{26h}),(\ref{26i}), (\ref{26j}), }           \nonumber \\
		&~{\rm (\ref{26d}), } \nonumber \\
		&~ \forall k \in  \mathcal{K},  m \in \mathcal{M},{\bf e}_k \in \mathcal{E}_k, {\bf e}_{\mathrm{Eve},m} \in \mathcal{E}_{\mathrm{Eve},m},{\bf e}_{\rm a} \in \mathcal{E}_{\rm a}, \nonumber
	\end{align}
\end{subequations}
where ${\bf L}_1 \triangleq {\bf F}\cup {\bf \Theta} \cup {\bf P}\cup {\bf q} \cup \{ \alpha_k, \alpha_{\rm a}\} \cup \{ {\bf W}_k, {\bf W}_{\rm a}, {\bf Z} \}$ and $\tau_1({\bf L}_1)$ is given by (\ref{27}). 
\begin{figure*}
	\begin{flalign}
		\tau_1({\bf L}_1 )  \triangleq E_I + \eta \left(  \sum_{k \in \mathcal{K}}{E_{u,k}[n]}+  \theta_{\rm a} \widehat{\Delta}_{\rm T} {\mathrm {Tr}}({\bf W}_{\rm a})	+ t_d \left(\sum_{k \in \mathcal{K}}{{\rm Tr}({\bf W}_k)}  +{\rm Tr}({\bf Z}) \right) + P(\Vert  {\bf v} \Vert) \Delta_{\rm T} \right) . \label{27}
	\end{flalign}
	\hrule
\end{figure*}
The channel errors involved in (\ref{26c}), (\ref{26i}), (\ref{26j}) and (\ref{26d}) bring infinite number of constraints in Problem ${\bf P}_2$, which makes Problem ${\bf P}_2$ computationally intractable. Define that
\begin{subnumcases}
	{}
	{\bar{\bf e}_k} = {\bf e}_k d_k \in \bar{\mathcal{E}}_k \triangleq \{ \bar{{\bf e}}_k | \bar{\bf e}_k^H {\bf C}_k \bar{\bf e}_k \leq d_k^2  \}, \nonumber \\
	{\bar{\bf e}_{{\rm Eve},m}} = {\bf e}_{{\rm Eve},m} d_{{\rm Eve},m} \in \bar{\mathcal{E}}_{{\rm Eve},m} \triangleq  \nonumber \\
	~~~~~~~~ \{ \bar{{\bf e}}_{{\rm Eve},m} | \bar{\bf e}_{{\rm Eve}, m}^H {\bf C}_{{\rm Eve},m} \bar{\bf e}_{{\rm Eve},m} \leq d_{{\rm Eve},m}^2  \},  \nonumber \\
	{\bar{\bf e}_{\rm a}} = {\bf e}_{\rm a} d_{\rm a} \in \bar{\mathcal{E}}_{\rm a} \triangleq \{ \bar{{\bf e}}_{\rm a} | \bar{\bf e}_{\rm a}^H {\bf C}_{\rm a} \bar{\bf e}_{\rm a} \leq d_{\rm a}^2  \}  \nonumber,
\end{subnumcases}
and then, constraints (\ref{26c}), (\ref{26i}), (\ref{26j}) and (\ref{26d}) are respectively rewritten as
\begin{subnumcases}
	{}
		\alpha_k d_k^2 \leq  \rho p_k N_{\mathrm T} + 2{\mathrm Re} \left\{ \bar{{\bf e}}_k^H \sqrt{\rho} p_k \bm {\mathrm a}_k \right\} + p_k \bar {\bf e}_k^H \bar {\bf e}_k,  \label{29a}\\
		 \sigma_{k}^2 d_k^2 + \rho \bm {\mathrm a}_k^H {\bf X}_{k,\mathrm{req}} \bm {\mathrm a}_k + 2{\mathrm{Re} \{\bar{\bf e}_k^H \sqrt{\rho} {\bf X}_{k,\mathrm{req}}  \bm {\mathrm a}_k \}}  \nonumber \\
		 + \bar {\bf e}_k^H {\bf X}_{k, \mathrm{req}} \bar {\bf e}_k  \leq 0, \label{29b}\\
		  \bar {\bf e}_{\mathrm{Eve},m}^H {\bf X}_{k,\mathrm{seq}} \bar {\bf e}_{\mathrm{Eve},m}+ 2{\mathrm{Re} \{ \bar {\bf e}_{\mathrm{Eve},m}^H \sqrt{\rho} {\bf X}_{k,\mathrm{seq}} \bm {\mathrm a}_{\mathrm{Eve},m} \}} \nonumber \\
		 +\sigma_{\mathrm{Eve}, m}^2 d_{\mathrm{Eve},m}^2  + \rho \bm {\mathrm a}_{\mathrm{Eve},m}^H {\bf X}_{k, \mathrm{seq}} \bm {\mathrm a}_{\mathrm{Eve},m}  \geq 0,  \label{29c}	\\
		 \alpha_{\rm a} d_{\rm a}^2 \leq \rho \bm {\mathrm a}_{\rm a}^H {\bf W}_{\rm a} \bm {\mathrm a}_{\rm a} + 2{\mathrm Re}\left\{\bar{\bf e}_{\rm a}^H \sqrt{\rho} {\bf W}_{\rm a} \bm {\mathrm a}_{\rm a}  \right\}  \nonumber \\
		 ~~~~~~~~  + \bar{\bf e}_{\rm a}^H {\bf W}_{\rm a} \bar{\bf e}_{\rm a}, \label{29d}		 
\end{subnumcases}
To handle the infinite number of constraints (\ref{29a}), (\ref{29b}), (\ref{29c}) and (\ref{29d}) due to the continuous channel errors, S-Procedure is employed, which is summarized as Lemma 1.
\begin{Lemma}{\rm \cite{xian12}}
	Let ${\bf F}_1 ,{\bf F}_2 \in \mathbb{H}^n $, ${\bf g}_1, {\bf g}_2 \in \mathbb{C}^n$, $h_1$, $h_2$ $\in \mathbb{R}$. The following implication
	\begin{align*}
		 {\bf x}^H {\bf F}_1 {\bf x} + 2 \mathrm{Re}\{ {\bf g}_1^H {\bf x}\} + h_1 \leq 0\\
		  \Rightarrow
		 {\bf x}^H {\bf F}_2 {\bf x} + 2 \mathrm{Re}\{{\bf g}_2^H {\bf x}\} + h_2 \leq 0
	\end{align*}
holds true if and only if there exists a $\lambda \geq 0$ such that
\begin{align*}
	\begin{bmatrix}
		{\bf F}_2 & {\bf g}_2 \\
		{\bf g}_2^H & h_2
	\end{bmatrix}	
	\preceq  \lambda
	\begin{bmatrix}
		{\bf F}_1 & {\bf g}_1\\
		{\bf g}_1^H & h_1
	\end{bmatrix}
\end{align*}
\end{Lemma}
\begin{IEEEproof}
	The proof of Lemma 1 is referred to \cite{xian12}, which is omitted here.
\end{IEEEproof}
With Lemma 1, constraints (\ref{29a}), (\ref{29b}), (\ref{29c}) and (\ref{29d}) are respectively equivalently expressed as
\begin{subnumcases}
	{}
	\begin{bmatrix}
		\lambda_k {\bf C}_k + p_k {\bf I}_{N_{\mathrm T}}&  p_k \sqrt{\rho}  {\bf a}_k \\ 	
		p_k\sqrt{\rho} {\bf a}_k^H & p_k\rho N_{\mathrm T} -(\alpha_k+\lambda_k)d_k^2
	\end{bmatrix}
\succeq {\bf 0},  \label{30a} \\
	\begin{bmatrix}
	\lambda_k^u {\bf C}_k -  {\bf X}_{k,\mathrm{req}}&  -\sqrt{\rho} {\bf X}_{k,\mathrm{req}}   {\bf a}_k \\ 	
	-\sqrt{\rho}  {\bf a}_k^H {\bf X}_{k,\mathrm{req}} &  -(\lambda_k^u+\sigma_{k}^2) d_k^2 - \rho  {\bf a}_k^H  {\bf X}_{k,\mathrm{req}}  {\bf a}_k
\end{bmatrix} \nonumber \\
~~~  \succeq {\bf 0},  \label{30b}\\
\begin{bmatrix}
	\lambda_{m,k}^u {\bf C}_{\mathrm{Eve},m} +  {\bf X}_{k,\mathrm{seq}}&  \sqrt{\rho} {\bf X}_{k,\mathrm{seq}}   {\bf a}_{\mathrm{Eve},m} \\ 	
	\sqrt{\rho} {\bf a}_{\mathrm{Eve},m}^H {\bf X}_{k,\mathrm{seq}} & \kappa
\end{bmatrix}
\succeq {\bf 0}, \label{30c} \\
 \begin{bmatrix}
	\lambda_{\rm a} {\bf C}_{\rm a} + {\bf W}_{\rm a} &  \sqrt{\rho}  {\bf W}_{\rm a}  {\bf a}_a \\ 	
	\sqrt{\rho} {\bf a}_{\rm a}^H {\bf W}_{\rm a}   & \rho {\bf a}_{\rm a}^H {\bf W}_{\rm a} {\bf a}_a -(\alpha_{\rm a}+\lambda_{\rm a})d_{\rm a}^2
\end{bmatrix}
\succeq {\bf 0},  \label{30d}
\end{subnumcases}
where $ \lambda_k $, $\lambda_{\rm a}$, $\lambda_k^u$ and $\lambda_{m,k}^u$ are non-negative auxiliary variables, and $\kappa$ is given by
\begin{flalign}
	\kappa = (\sigma_{\mathrm{Eve},m}^2-\lambda_{m,k}^u) d_{\mathrm{Eve},m}^2 + \rho \bm {\mathrm a}_{\mathrm{Eve},m}^H  {\bf X}_{k,\mathrm{seq}} \bm {\mathrm a}_{\mathrm{Eve},m}.
\end{flalign}
However, constraints (\ref{30a}), (\ref{30b}), (\ref{30c}) and (\ref{30d}) are still non-convex due to the coupling variables. The following Proposition 1 is presented to handle them.
\begin{mypro}
	Let ${\bf A} \in \mathbb{H}^n$, ${\bf b} \in \mathbb{C}^n$, $c,d \in \mathbb{R}$. Then, the following implication
\begin{align*}
	\begin{bmatrix}
		{\bf A} & {\bf b} \\
		{\bf b}^H & c
	\end{bmatrix}
\succeq \bm 0
\Rightarrow
	\begin{bmatrix}
	{\bf A} & {\bf b} \\
	{\bf b}^H & d
\end{bmatrix}
\succeq \bm 0
\end{align*}
holds true if $d \geq c$.
\end{mypro}
\begin{IEEEproof}
	Define ${\bf F}, {\bf G} $ and $ {\bf H} $ as follows
	\begin{align*}
		{\bf F} =
		\begin{bmatrix}
			{\bf A} & {\bf b} \\
			{\bf b}^H & c
		\end{bmatrix},
	{\bf G} = \begin{bmatrix}
		{\bf A} & {\bf b} \\
		{\bf b}^H & d
	\end{bmatrix},
{\bf H} = \begin{bmatrix}
	\bm 0 & \bm 0 \\
	\bm 0 & d-c
\end{bmatrix}.
	\end{align*}
One can find that ${\bf G} = {\bf F} + {\bf H}$ and ${\bf H }\succeq \bm 0 $ with $d \geq c$. Then ${\bf G} \succeq \bm 0$ holds true if ${\bf F} \succeq \bm 0$.
\end{IEEEproof}
Define auxiliary variables $\beta_k, \beta_a, \beta_k^u$ and $\beta_{m,k}^u$ satisfying that
\begin{subnumcases}
	{}
		\beta_k \geq  (\alpha_k+\lambda_k)d_k^2, \label{31a}\\
		\beta_{\rm a} \geq  (\alpha_{\rm a}+\lambda_{\rm a})d_{\rm a}^2,  \label{31b}\\
		\beta_k^u \geq (\lambda_k^u+\sigma_{k}^2) d_k^2,  \label{31c} \\
		\beta_{m,k}^u \leq (\sigma_{\mathrm{Eve},m}^2-\lambda_{m,k}^u) d_{\mathrm{Eve},m}^2, \label{31d}
\end{subnumcases}
and according to Proposition 1, constraints (\ref{30a}), (\ref{30b}), (\ref{30c}) and (\ref{30d}) are respectively expressed as the following convex constraints, i.e.,
\begin{subequations}
	\begin{align}
		&\begin{bmatrix}
			\lambda_k{\bf C}_k + p_k {\bf I}_{N_{\mathrm T}}&  p_k \sqrt{\rho} \bm {\mathrm a}_k \\ 	
			p_k\sqrt{\rho}\bm {\mathrm a}_k^H & p_k\rho N_{\mathrm T} -\beta_k
		\end{bmatrix}
		\succeq \bm 0, \label{32a} \\
		&\begin{bmatrix}
			\lambda_k^u {\bf C}_k -  {\bf X}_{k,\mathrm{req}}&  -\sqrt{\rho} {\bf X}_{k,\mathrm{req}}  \bm {\mathrm a}_k \\ 	
			-\sqrt{\rho} \bm {\mathrm a}_k^H {\bf X}_{k,\mathrm{req}} &  -\beta_k^u - \rho \bm {\mathrm a}_k^H  {\bf X}_{k,\mathrm{req}} \bm {\mathrm a}_k
		\end{bmatrix}  \succeq \bm 0, \label{32b}\\
		&\begin{bmatrix}
			\lambda_{m,k}^u {\bf C}_{\mathrm{Eve},m} +  {\bf X}_{k,\mathrm{seq}}&  \sqrt{\rho} {\bf X}_{k,\mathrm{seq}}  \bm {\mathrm a}_{\mathrm{Eve},m} \\ 	
			\sqrt{\rho} \bm {\mathrm a}_{\mathrm{Eve},m }^H {\bf X}_{k,\mathrm{seq}} & \beta_{m,k}^u + \rho \bm {\mathrm a}_{\mathrm{Eve}, m}^H  {\bf X}_{k,\mathrm{seq}} \bm {\mathrm a}_{\mathrm{Eve},m}
		\end{bmatrix}
		\nonumber \\
		&~\succeq \bm 0,  \label{32c} \\
		&{\rm and}~~~\begin{bmatrix}
			\lambda_{\rm a} {\bf C}_{\rm a} + {\bf W}_{\rm a} &  \sqrt{\rho}  {\bf W}_{\rm a} \bm {\mathrm a}_a \\ 	
			\sqrt{\rho}\bm {\mathrm a}^H {\bf W}_{\rm a}   & \rho \bm {\mathrm a}_{\rm a}^H {\bf W}_{\rm a} \bm {\mathrm a}_{\rm a} -\beta_{\rm a}
		\end{bmatrix}
		\succeq \bm 0 \label{32d}. 
	\end{align}	
\end{subequations}
Then, Problem \textbf{P$_2$} is equivalently expressed as Problem \textbf{P$_3$}.
\begin{subequations}
	\begin{align}
		\textbf{P$_3$}:&\mathop{\min}\limits_{{\bf L}_2}
		~~\tau_1\left( {\bf L}_2 \right)  \label{33a}~~~~~\\ 
		{\rm s.t.}&~ \lambda_k \geq 0, \lambda_a \geq 0, \lambda_k^u \geq 0, \lambda_{m,k}^u \geq 0,                    \label{33c}\\	
		&~{\rm (\ref{shi1}),(\ref{shi8}),(\ref{shi19}),(\ref{23b}), (\ref{23c}),(\ref{23d}),(\ref{23i}),(\ref{23j}),(\ref{23k})}, \nonumber \\
		&~ {\rm (\ref{23l}),(\ref{25a}),(\ref{25b}),(\ref{26g}),(\ref{26h}),(\ref{26e}),(\ref{26f}), (\ref{31a}),} \nonumber \\
		&~{\rm (\ref{31b}), (\ref{31c}), (\ref{31d}), (\ref{32a}), (\ref{32b}), (\ref{32c}),(\ref{32d}),}                    \nonumber \\
		&~\forall k \in \mathcal{K}, m \in \mathcal{M}, \nonumber
	\end{align}
\end{subequations}
where ${\bf L}_2 \triangleq {\bf L}_1 \cup \{ \lambda_k, \lambda_{\rm a}, \lambda_k^u, \lambda_{m,k}^u \} \cup \{ \beta_k, \beta_{\rm a}, \beta_k^u, \beta_{m,k}^u \}$. Although the computationally intractable issue due to the infinite number of channel errors is addressed, the objective function (\ref{33a}) and constraints  (\ref{shi19}), (\ref{25a}), (\ref{25b}), (\ref{31a}), (\ref{31b}), (\ref{31c}) and (\ref{31d}) in Problem ${\bf P}_3$ are still non-convex. \\

\indent
It is observed that $\theta_k, \theta_{\rm a}$ and $f_{u,k}$ are non-negative, which thus, can be substituted by $e^{t_k}, e^{t_{\rm a}}$ and $e^{g_k}$, respectively. (\ref{shi8}), (\ref{shi19}), (\ref{23c}), (\ref{25a}) and (\ref{25b}) are respectively re-expressed as
\begin{subnumcases}
	{}
		~	\sum\nolimits_{k \in \mathcal{K}}{e^{t_k}} + e^{t_{\rm a}} \leq 1, \label{34d} \\
		~Q_k \geq \frac{f_{{\rm l},k}}{D_k} \widehat{\Delta}_{\rm T} + \frac{e^{g_k}}{D_k} e^{t_{\rm a}} \widehat{\Delta}_{\rm T},\label{34c} \\
		~ 	\sum\nolimits_{k \in \mathcal{K}}{e^{g_k}} \leq f_{{\rm u}, \mathrm{max}}, \label{34f}	\\
		~ e^{t_k} \widehat{\Delta}_{\rm T} B \log_2 \left( {1+\frac{\alpha_k}{\sigma_k^2}} \right)  \geq Q_k - \frac{f_{{\rm l},k}}{D_k} \widehat{\Delta}_{\rm T} , \label{34g} \\
		~ e^{t_{\rm a}} B \log_2 \left( {1+\frac{ \alpha_{\rm a}}{\sigma_{\rm a}^2}} \right)  \geq \nonumber \\
		~~~~~~ \sum_{k \in \mathcal{K}}{\left( \frac{Q_k}{ \widehat{\Delta}_{\rm T}}-\frac{f_{{\rm l},k}}{D_k}  - \frac{e^{g_k+t_{\rm a}}}{D_k} \right)},  \label{34h} 
\end{subnumcases}

Then, Problem \textbf{P$_3$} is rewritten as
\begin{subequations}
	\begin{align}
		\textbf{P$_4$}:&\mathop{\min}\limits_{{\bf L}_3}
		~~\tau_2\left( {\bf L}_3 \right)  \label{34a}~~~~~\\
		{\rm s.t.}&~{\rm (\ref{shi1}),(\ref{23b}),(\ref{23d}),(\ref{23i}),(\ref{23j}),(\ref{23l}), (\ref{26g}),(\ref{26h}),(\ref{26e}),} \nonumber \\
		&~{\rm (\ref{26f}),(\ref{31a}),(\ref{31b}),(\ref{31c}),(\ref{31d}), (\ref{32a}),(\ref{32b}),(\ref{32c}), }               \nonumber \\
		&~{\rm  (\ref{32d}), (\ref{33c}), (\ref{34d}), (\ref{34c}),(\ref{34f}),(\ref{34g}),(\ref{34h}),}
		\nonumber\\
		&~\forall k \in \mathcal{K}, m \in \mathcal{M}, \nonumber
	\end{align}
\end{subequations}
where ${\bf L}_3 \triangleq {\bf L}_2 \setminus \{ \theta_k, \theta_{\rm a}, f_{{\rm u},k} \} \cup \{ t_k, t_{\rm a}, g_k\} $ and $\tau_2({\bf L}_3)$ is given by (\ref{35}).
\begin{figure*}
	\begin{flalign}
		\tau_2({\bf L}_3) \triangleq &~ \eta \left(
		\sum_{k \in \mathcal{K}} v_{\rm u} e^{3g_k+t_{\rm a}} \widehat{\Delta}_{\rm T}
		+  e^{t_{\rm a}} \widehat{\Delta}_{\rm T} {\mathrm {Tr}}({\bf W}_{\rm a})+
		t_{\rm d} \left(\sum_{k \in \mathcal{K}}{{\rm Tr}({\bf W}_k)}  +{\rm Tr}({\bf Z}) \right)+
		 P(\Vert  {\bf v} \Vert) \Delta_{\rm T}
		\right) \nonumber \\
		&~ + \sum_{k \in \mathcal{K}}\left(
		v_{\rm l} f_{{\rm l},k}^3 \widehat{\Delta}_{\rm T} +
		e^{t_k} \widehat{\Delta}_{\rm T} p_k
		\right) .  
		\label{35}
	\end{flalign}
	\hrule
\end{figure*}
It is observed that Problem \textbf{P$_4$} is also non-convex due to the objective function (\ref{34a}) and the constraints (\ref{31a}), (\ref{31b}), (\ref{31c}), (\ref{31d}), (\ref{34g}) and (\ref{34h}). To handle  the non-convex constraints (\ref{31a}), (\ref{31b}), (\ref{31c}) and (\ref{31d}), define auxiliary variables $\gamma_k$, $\gamma_{\rm a}$, $\gamma_k^u$, $\gamma_{m,k}^u$, $s_k$, $s_{\mathrm {Eve},m}$ and $s_{\rm a}$ satisfying that,

	\begin{flalign}
		\left\{ \begin{array}{l}
			{\alpha _k} + {\lambda _k} \le {e^{{\gamma _k}}}, \\
			{\alpha _{\rm{a}}} + {\lambda _{\rm{a}}} \le {e^{{\gamma _{\rm{a}}}}},\\
			\lambda _k^u \le {e^{\gamma _k^u}}, \\
			\lambda _{m,k}^u \le {e^{\gamma _{m,k}^u}}, \\
			{\left\| {{\bf{q}} - {{\bf{u}}_k}} \right\|^2} + {H^2} \le {e^{{s_k}}},\\
			{\left\| {{\bf{q}} - {{\bf{u}}_{{\rm{Eve}},m}}} \right\|^2} + {H^2} \le {e^{{s_{{\rm{Eve}},m}}}},\\
			{\left\| {{\bf{q}} - {{\bf{u}}_{\rm{a}}}} \right\|^2} + {H^2} \le {e^{{s_{\rm{a}}}}}.
		\end{array} \right. \label{39}
	\end{flalign}

Then, constraints (\ref{31a}), (\ref{31b}), (\ref{31c}) and (\ref{31d}) are respectively expressed as
\begin{subnumcases}
	{}
		\beta_k \geq e^{\gamma_k+s_k},\label{40a}   \\
		\beta_{\rm a} \geq e^{\gamma_{\rm a}+s_{\rm a}}, \label{40b} \\
		\beta_k^u \geq e^{\gamma_k^u+s_k} + \sigma_{k}^2({\Vert {\bf q}-{\bf u}_k \Vert}^2 + H^2 ),   \label{40c} ~~~~~~~   \\
		\beta_{m,k}^u + e^{\gamma_{m,k}^u+s_{\mathrm{Eve},m}} \leq \nonumber \\
		~~~~~~\sigma_{\mathrm{Eve},m}^2({\Vert {\bf q}-{\bf u}_{\mathrm{Eve},m} \Vert}^2 + H^2 ). \label{40d}
\end{subnumcases}

To handle the non-convex constraints (\ref{34g}) and (\ref{34h}), define auxiliary variables $r_k$ and $r_{\rm a}$ satisfying that,
\begin{flalign}
	\left\{ \begin{array}{l}
		{e^{{r_k}}} \le {\log _2}\left( {1 + \frac{{{\alpha _k}}}{{\sigma _k^2}}} \right),\\
		{e^{{r_{\rm{a}}}}} \le {\log _2}\left( {1 + \frac{{{\alpha _{\rm{a}}}}}{{\sigma _{\rm{a}}^2}}} \right).
	\end{array} \right.  \label{37}
\end{flalign}

Then, constraints (\ref{34g}) and (\ref{34h}) are respectively expressed as
\begin{flalign}
	\left\{ \begin{array}{l}
		{e^{{t_k} + {r_k}}}{{\widehat \Delta }_{\rm{T}}}B \ge {Q_k} - \frac{{{f_{{\rm{l}},k}}}}{{{D_k}}}{{\widehat \Delta }_{\rm{T}}},\\
		{e^{{t_{\rm{a}}} + {r_{\rm{a}}}}}B + \sum\limits_{k \in {\cal K}} {\frac{{{e^{{g_k} + {t_{\rm{a}}}}}}}{{{D_k}}}}  \ge \sum\nolimits_{k \in {\cal K}} {(\frac{{{Q_k}}}{{{{\widehat \Delta }_{\rm{T}}}}} - \frac{{{f_{{\rm{l}},k}}}}{{{D_k}}})}. 
	\end{array} \right. \label{38}
\end{flalign}
 Then, Problem \textbf{P$_4$} is rewritten as
 \begin{subequations}
 	\begin{align}
 		\textbf{P$_5$}:&\mathop{\min}\limits_{{\bf L}_4}
 		~~\tau_2\left( {\bf L}_4 \right)  \label{41a}~~~~~\\
 		{\rm s.t.}&~{\rm (\ref{shi1}),(\ref{23b}),(\ref{23d}),(\ref{23i}),(\ref{23j}),(\ref{23l}), (\ref{26g}),(\ref{26h}),(\ref{26e}),} \nonumber \\
 		&~{\rm (\ref{26f}),(\ref{32a}),(\ref{32b}),(\ref{31c}),(\ref{32d}), (\ref{33c}), (\ref{34c}),(\ref{34f})},\nonumber \\
 		&~{\rm (\ref{34d}),(\ref{39}),(\ref{40a}),(\ref{40b}),(\ref{40c}),(\ref{40d}),(\ref{37}),(\ref{38})} ,   \nonumber \\
 		& \forall k \in \mathcal{K}, m \in \mathcal{M}, \nonumber
 	\end{align}
 \end{subequations}
where ${\bf L}_4 \triangleq {\bf L}_3 \cup \{ r_k, r_{\rm a}, \gamma_k, \gamma_{\rm a}, \gamma_k^u, \gamma_{m,k}^u, s_k, s_{\mathrm{Eve},m }, s_{\rm a} \}$. Problem ${\bf P}_5$ is non-convex due to the non-convex objective function and the constraints (\ref{39}), (\ref{40d}) and (\ref{38}). We first rewrite the objective function into a convex form and then, transform the non-convex constraints into convex ones by first-order approximation. Define auxiliary variable $\zeta_k, \zeta_{\rm a}, v_1$ and $v_2$ satisfying that
\begin{subnumcases}
	{}
		e^{\zeta_k} \geq p_k, \forall k \in \mathcal{K}, \label{42a} \\
		e^{\zeta_{\rm a}} \geq {\mathrm {Tr}}({\bf W}_{\rm a}), \label{42b} \\
		v_1\geq \frac{\Vert {\bf q}[n] - {\bf q}[n-1] \Vert}{\Delta_{\rm T}}, \label{42c} \\
		v_2^2 \geq \sqrt{1 +\frac{{\Vert {\bf q}[n] - {\bf q}[n-1] \Vert}^4}{4 v_0^4 \Delta_{\rm T}^4}}
		\nonumber \\
		~~~~~~~-\frac{  {\Vert {\bf q}[n] - {\bf q}[n-1] \Vert}^2}{2 v_0^2 \Delta_{\rm T}^2}, \label{42d}
\end{subnumcases}
and (\ref{42d}) is further re-expressed as 
\begin{flalign}
	v_2^2 + \frac{{\Vert {\bf q}[n] - {\bf q}[n-1] \Vert}^2}{v_0^2 \Delta_{\rm T}^2}  \geq \frac{1}{v_2^2}, \label{43} 
\end{flalign}

Then, Problem \textbf{P$_5$} is rewritten as
 \begin{subequations}
	\begin{align}
		\textbf{P$_6$}:&\mathop{\min}\limits_{{\bf L}_5}
		~~\tau_3\left( {\bf L}_5 \right)  \label{44a}~~~~~\\
		{\rm s.t.}&~{\rm (\ref{shi1}),(\ref{23b}),(\ref{23d}),(\ref{23i}),(\ref{23j}),(\ref{23l}), (\ref{26g}),(\ref{26h}),(\ref{26e}),} \nonumber \\
		&~{\rm (\ref{26f}),(\ref{32a}),(\ref{32b}),(\ref{32c}),(\ref{32d}), (\ref{33c}),(\ref{34d}), (\ref{34c}),} \nonumber \\
		&~{\rm (\ref{34f}),(\ref{39}),(\ref{40a}),(\ref{40b}),(\ref{40c}),(\ref{40d}), (\ref{37}), (\ref{38}),(\ref{42a}),} \nonumber \\
	    &~{\rm (\ref{42b}),(\ref{42c}),(\ref{43}),}              \nonumber \\
	    & \forall k \in \mathcal{K}, m \in \mathcal{M}, \nonumber
	\end{align}
\end{subequations}
where ${\bf L}_5 \triangleq {\bf L}_4 \cup \{ \zeta_k, \zeta_{\rm a}, v_1, v_2 \}$ and $\tau_3({\bf L}_5)$ is given by (\ref{45}).
\begin{figure*}
	\begin{flalign}
		\tau_3({\bf L}_5) \triangleq  &\eta \left(
		\sum_{k \in \mathcal{K}} v_{\rm u} e^{3g_k+t_{\rm a}} \widehat{\Delta}_{\rm T}
		+  e^{t_{\rm a}+\zeta_{\rm a}} \widehat{\Delta}_{\rm T} +
		t_{\rm d} \left(\sum_{k \in \mathcal{K}}{{\rm Tr}({\bf W}_k)}  +{\rm Tr}({\bf Z}) \right)+
				\left(P_0 \left(1 + \frac{3v_1^2}{U_{\rm tip}^2}\right)
				+ \frac{1}{2} d_0 \rho_0 s A v_1^3
				+ P_{\rm H} v_2\right) \Delta_{\rm T}
		\right) \nonumber \\
		& + \sum_{k \in \mathcal{K}}\left(
		v_{\rm l} f_{{\rm l},k}^3 \widehat{\Delta}_{\rm T} +
		e^{t_k+\zeta_k} \widehat{\Delta}_{\rm T}
		\right).
		\label{45}
	\end{flalign}
	\hrule
\end{figure*}
The objective function of Problem $ {\bf P}_6$ is convex, while the constraints (\ref{39}), (\ref{38}), (\ref{40d}), (\ref{42a}), (\ref{42b}) and (\ref{43}),  are still non-convex. Nevertheless, both sides of the above constraints are convex functions. Via first-order approximation, (\ref{39}), (\ref{38}), (\ref{40d}), (\ref{42a}), (\ref{42b}) and (\ref{43}) are respectively approximated by the following convex constraints.
\begin{flalign}
	\left\{ \begin{array}{l}
		{\alpha _k} + {\lambda _k} \le {e^{{{\bar \gamma }_k}}} + {e^{{{\bar \gamma }_k}}}({\gamma _k} - {{\bar \gamma }_k}),\\
		{\alpha _{\rm{a}}} + {\lambda _{\rm{a}}} \le {e^{{{\bar \gamma }_{\rm{a}}}}} + {e^{{{\bar \gamma }_{\rm{a}}}}}({\gamma _{\rm{a}}} - {{\bar \gamma }_{\rm{a}}}),\\
		\lambda _k^u \le {e^{\bar \gamma _k^u}} + {e^{\bar \gamma _k^u}}(\gamma _k^u - \bar \gamma _k^u),\\
		\lambda _{m,k}^u \le {e^{\bar \gamma _{m,k}^u}} + {e^{\bar \gamma _{m,k}^u}}(\gamma _{m,k}^u - \bar \gamma _{m,k}^u),\\
		{\left\| {{\bf{q}} - {{\bf{u}}_k}} \right\|^2} + {H^2} \le {e^{{{\bar s}_k}}} + {e^{{{\bar s}_k}}}({s_k} - {{\bar s}_k}),\\
		{\left\| {{\bf{q}} - {{\bf{u}}_{{\rm{Eve}},m}}} \right\|^2} + {H^2} \le \chi_1({s_{{\rm Eve},m}}),\\
		{\left\| {{\bf{q}} - {{\bf{u}}_{\rm{a}}}} \right\|^2} + {H^2} \le {e^{{{\bar s}_{\rm{a}}}}} + {e^{{{\bar s}_{\rm{a}}}}}({s_{\rm{a}}} - {{\bar s}_{\rm{a}}}),\\
		{\chi _2}({t_k},{r_k}){{\widehat \Delta }_{\rm{T}}}B \ge {Q_k} - \frac{{{f_{{\rm{l}},k}}}}{{{D_k}}}{{\widehat \Delta }_{\rm{T}}},\\
		{\chi _3}({t_a},{r_a},{g_k}) \ge \sum\nolimits_{k \in {\cal K}} {\left( {\frac{{{Q_k}}}{{{{\widehat \Delta }_{\rm{T}}}}} - \frac{{{f_{{\rm{l}},k}}}}{{{D_k}}}} \right)}, \\
		\beta _{m,k}^u + {e^{\gamma _{m,k}^u + {s_{{\rm{Eve}},m}}}} \le \sigma _{{\rm{Eve}},m}^2\chi _4^m({\bf{q}}),\\
		{e^{{{\bar \zeta }_k}}} + {e^{{{\bar \zeta }_k}}}({\zeta _k} - {{\bar \zeta }_k}) \ge {p_k},\\
		{e^{{{\bar \zeta }_{\rm{a}}}}} + {e^{{{\bar \zeta }_{\rm{a}}}}}({\zeta _{\rm{a}}} - {{\bar \zeta }_{\rm{a}}}) \ge {\rm{Tr}}({{\bf{W}}_{\rm{a}}}),\\
		{\chi _5}({\bf{q}}[n],{v_2}) \ge \frac{1}{{v_2^2}},
	\end{array} \right. \label{46}
\end{flalign}
where $ \chi_1({s_{{\rm Eve},m}}) $, $\chi_2(t_k,r_k)$, $\chi_3(t_a,r_a,g_k)$, $\chi_4^m({\bf q})$  and $ \chi_5({\bf q}[n], v_2)$ are represented in (\ref{47}). To improve the approximation precise, the SCA method is employed. In particular, in each iteration, the following convex Problem ${\bf P}_7$ is required to be solved with a feasible point ${\bar {\bf L}}_5$. 
\begin{figure*}
\begin{flalign}
\left\{ \begin{array}{l}
	{\chi _1}({s_{{\rm{Eve}},m}}) \triangleq {e^{{{\bar s}_{{\rm{Eve}},m}}}} + {e^{{{\bar s}_{{\rm{Eve}},m}}}}({s_{{\rm{Eve}},m}} - {{\bar s}_{{\rm{Eve}},m}}),\\
	{\chi _2}({t_k},{r_k}) \triangleq {e^{{{\bar t}_k} + {{\bar r}_k}}} + {e^{{{\bar t}_k} + {{\bar r}_k}}}({t_k} + {r_k} - {{\bar t}_k} - {{\bar r}_k}),\\
	{\chi _3}({t_a},{r_a},{g_k}) \triangleq {e^{{{\bar t}_{\rm{a}}} + {{\bar r}_{\rm{a}}}}}B + {e^{{{\bar t}_{\rm{a}}} + {{\bar r}_{\rm{a}}}}}({t_{\rm{a}}} + {r_{\rm{a}}} - {{\bar t}_{\rm{a}}} - {{\bar r}_{\rm{a}}})B + \sum\limits_{k \in {\cal K}} {\frac{{{e^{{{\bar g}_k} + {{\bar t}_{\rm{a}}}}} + {e^{{{\bar g}_k} + {{\bar t}_{\rm{a}}}}}({g_k} + {t_{\rm{a}}} - {{\bar g}_k} - {{\bar t}_{\rm{a}}})}}{{{D_k}}}}, \\
	\chi _4^m({\bf{q}}) \triangleq {\left\| {\overline {\bf{q}}  - {{\bf{u}}_{{\rm{Eve}},m}}} \right\|^2} + {H^2} + 2{(\overline {\bf{q}}  - {{\bf{u}}_{{\rm{Eve}},m}})^T}({\bf{q}} - \overline {\bf{q}} ),\\
	{\chi _5}({\bf{q}}[n],{v_2}) \triangleq \bar v_2^2 + \frac{{{{\left\| {\overline {\bf{q}} [n] - {\bf{q}}[n - 1]} \right\|}^2}}}{{v_0^2\Delta _{\rm{T}}^2}} + 2{{\bar v}_2}({v_2} - {{\bar v}_2}) + \frac{{2{{(\overline {\bf{q}} [n] - {\bf{q}}[n - 1])}^T}({\bf{q}}[n] - \overline {\bf{q}} [n])}}{{v_0^2\Delta _{\rm{T}}^2}}.
\end{array} \right. \label{47}
\end{flalign}
\hrule
\end{figure*}

 \begin{subequations}
	\begin{align}
		\textbf{P$_7$}:&\mathop{\min}\limits_{{\bf L}_5}
		~~\tau_3\left( {\bf L}_5 \right)  \label{48a}~~~~~\\
		{\rm s.t.}&~{\rm (\ref{shi1}),(\ref{23b}),(\ref{23d}),(\ref{23i}),(\ref{23j}),(\ref{23l}), (\ref{26g}),(\ref{26h}),(\ref{26e}),} \nonumber \\
		&~{\rm (\ref{26f}),(\ref{32a}),(\ref{32b}),(\ref{32c}),(\ref{32d}), (\ref{33c}), (\ref{34d}),(\ref{34c}),} \nonumber \\
		&~{\rm (\ref{34f}),(\ref{40a}),(\ref{40b}),(\ref{40c}), (\ref{37}), (\ref{42c}),(\ref{46}),}                \nonumber\\
		&~\forall k \in \mathcal{K}, m \in \mathcal{M}. \nonumber
	\end{align}
\end{subequations}
\noindent
The proposed algorithm for solving Problem ${\bf P}_1$ is summarized in Algorithm 1.
\begin{algorithm}[h]
	\caption{The proposed algorithm for solving Problem \textbf{P$_1$}} \label{alg:3}
	{\textbf{Setting}}: \\
	~Given the time slot index $n$ and tolerance error $\epsilon$; \\
	{\textbf{Initialization}: }\\
	~Calculate the channel vectors $\bm {\mathrm a}_k$, $\bm {\mathrm a}_{\mathrm{Eve},m}$, $\bm {\mathrm a}_{\mathrm a}$ through (\ref{shi 1a})-(\ref{shi 1c}) with ${\bf q}[n-1]$;\\
	~Set iteration step $i = 0$;
	\\
	~Obtain a feasible point $\bar{{\bf L}}_5[i]$ to Problem \textbf{P$_6$}; \\
	~Set ${\bf L}_5^{\star}[i] = \bar{{\bf L}}_5[i]$;\\
	\While{~\rm 1 }{
		~Update $i = i+1$;		\\					
		~Obtain the optimal solution ${\bf L}_5^{\star}[i]$ through solving Problem \textbf{P$_7$} with $\bar {\bf L}_5[i-1]$;\\
		~{\bf If} $\vert \tau_3({\bf L}_5^{\star}[i])- \tau_3({\bf L}_5^{\star}[i-1])\vert \leq \epsilon$: ~{\bf break};\\
		~Update $\bar{\bf L}_5[i] = {\bf L}_5^{\star}[i]$;\\
	}	
\end{algorithm}\\
\begin{mypro}
	The energy consumption sequence $\{\tau_3({\bf L}_5^{\star}[i]) \}_{i=1}^{\infty}$ generated by Algorithm 1 converges as the iteration step goes infinity.
\end{mypro}
\begin{IEEEproof}
	It is observed that the optimal solution ${\bf L}_5^{\star}[i-1]$ to Problem \textbf{P$_7$} in the $(i-1)$th iteration is a feasible solution to Problem \textbf{P$_7$} in the $i$th iteration. That is, it holds that $\tau_3({\bf L}_5^{\star}[i]) \leq \tau_3({\bf L}_5^{\star}[i-1])$, which indicates that $\{\tau_3({\bf L}_5^{\star}[i]) \}_{i=1}^{\infty}$ is  monotonically decreased  by Algorithm 1. Besides, as $\{\tau_3({\bf L}_5^{\star}[i]) \}_{i=1}^{\infty}$ denotes the energy consumption, it holds that $\{\tau_3({\bf L}_5^{\star}[i]) \}_{i=1}^{\infty} \geq 0$, which indicates that $\{\tau_3({\bf L}_5^{\star}[i]) \}_{i=1}^{\infty}$ is bounded. As a results, Proposition 2 is proved. 
\end{IEEEproof}
\indent
The complexity of the proposed algorithm is analyzed as follows. It is observed that in Algorithm 1, Problem \textbf{P}$_7$ is iteratively solved, which comprises of linear matrix inequalities (LMIs) and second-order cone (SOC) constraints. By solving Problem \textbf{P}$_7$ via a standard interior point method \cite{alg1}, the computation complexity is analyzed as follows. The decision number in Problem \textbf{P}$_7$ is $\iota = ( (K+2)N_{\rm T}^2 + 3MK+ 14K + M+ 12)$. By ignoring the low-order-complexity constraints in Problem \textbf{P}$_7$ \cite{alg2}, e.g., (\ref{23b}), (\ref{23d}) and (\ref{26g}), there are $3MK$ LMIs with the size of $1$, $K$ LMIs with the size of $N_{\rm T}$, $MK$ LMIs with the size of $(N_{\rm T}+1)$ and $(2K+M)$ SOCs with the size of $2$. Then, the computational complexity for solving the Problem \textbf{P}$_7$ is given in (\ref{com}). Assuming that Algorithm 1 requires $I$ iterations to converge, the total computational complexity for solving Problem \textbf{P}$_1$ is $I\mathcal{O} \left\{
[MKN_{\rm T}]^{1/2} \iota [\iota MKN_{\rm T}^3+ \iota^2 MK N_{\rm T}^2]
\right\}.$
\begin{figure*}
\begin{flalign}
	\mathcal{O} & \left\{ 
	\left[ 3MK + K N_{\rm T} + MK (N_T+1)+ 2(2K+M)  \right]^{1/2} \iota 
	\left[ \iota \left( K N_{\rm T}^3 + MK(N_{\rm T}+1)^3+ 3MK \right) + \iota^2\left( K N_{\rm T}^2  +MK(N_{\rm T}+1)^2+ \right. \right. \right. \nonumber \\
	&\left. \left. \left.
	 3MK \right) + 4\iota \left( 2K+M \right)
	   \right]
 	\right\}  = 
 	\mathcal{O} \left\{
 	[MKN_{\rm T}]^{1/2} \iota [\iota MKN_{\rm T}^3+ \iota^2 MK N_{\rm T}^2]
 	\right\}
 	\label{com}
\end{flalign}
\hrule
\end{figure*}

\section{Numerical Results}
\indent In this section, numerical results are presented to validate converge behavior of the proposed algorithm as well as deriving insights for network design in various scenarios compared with some existing benchmarks. The number of transmit antenna deployed at the UAV is set as $4$. The numbers of the GUs and eavesdroppers are set as $6$ and $2$, respectively. The horizontal coordinates of the GUs, eavesdroppers and ground MEC are set as ${\bf u}_1 = (0,20)$, ${\bf u}_2 = (10,10)$, ${\bf u}_3 = (15,40)$, ${\bf u}_4 = (40,20)$, ${\bf u}_5 = (30, 10)$, ${\bf u}_6 = (40, 20)$, ${\bf u}_{{\rm Eve},1} = (10, 0)$, ${\bf u}_{{\rm Eve}, 2}=(30,0)$ and ${\bf u}_{\rm a}=(20, 20)$, respectively. Other system parameters are shown in Table I unless otherwise specified.
%
%
\begin{table}[!t]
	\caption{INSTANCE PARAMETERS}
	\centering
	\begin{tabular}{cc|cc}
			\hline
			{\bf Parameters}& {\bf Values}&	{\bf Parameters}& {\bf Values}\\
			\hline
			$T$ & 6 s & 	$t_{\rm d}$ & 0.003 s \\

			$H$ & 20 m & $D_k$ & 10$^3$ cycles/bit\\

			$N$ & 20 & $B$ & 30 MHz  \\

			 ${\bf q}_{\rm I}$ & (0, 0) &$\sigma_k^2$, $\sigma_{{\rm Eve},m}^2$ $\sigma_{\rm a}^2$ & 10$^{-8} $ W\\

			 ${\bf q}_{\rm F}$ & (0, 40) &$I_k$& 10 Mbits \\

            $P_{\rm 0}$ & 225 W &$v_{\rm l}$, $v_{\rm u}$ & 10$^{-26}$\\
			
			$P_{\rm H}$ & 426 W&$f_{\rm l,max}$ &  0.3 GHz \\
			
			$U_{\rm tip}$ & 120 m/s & $f_{\rm u, max}$ & 6 GHZ\\
			
			$\rho_0$ & 1.225 kg/m$^3$ & $p_{\rm l, max}$ & 8 W\\ 
			
			$A$ & 0.503 m$^2$ & $p_{\rm u, max}$& 20 W\\
			
			$d_0$ & 0.6 & $\Gamma_{\rm req}$ & -10 dB\\
			
			$s$ & 0.05 & $\Gamma_{\rm seq}$ & -15 dB \\
			
			${\bf C}_k, {\bf C}_{{\rm Eve},m}, {\bf C}_{\rm a}$ & $10^{10} {\bf I}_{N_{\rm T}}$ & $V_{\max}$ & 20 m/s \\ 
			$\eta$ & 0.01  \\ \hline 
		\end{tabular}
\end{table}

\begin{figure}[t]
	\begin{center}
		\centerline{\includegraphics[ width=.5\textwidth]{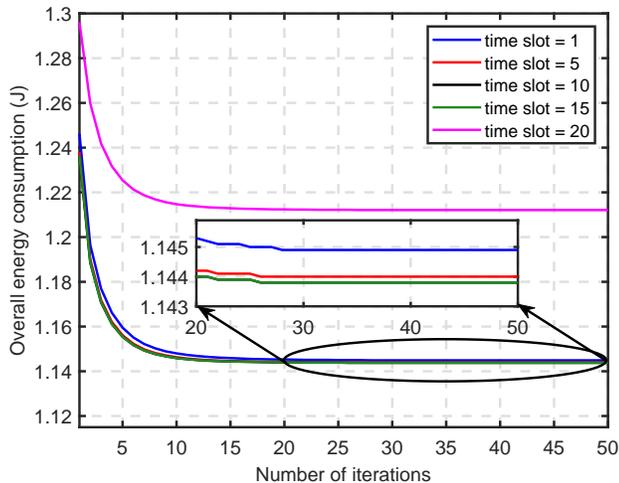}}
		\caption{The convergence behaviour of the proposed algorithm.}
		\label{fig:uav_con}
	\end{center}
\end{figure}

\indent Fig. \ref{fig:uav_con} shows the convergence behavior of the proposed algorithm for time slots of $1$, $5$, $10$, $15$ and $20$. It is observed that for all time slots, Algorithm 1 converges, which is consistent with Proposition 2. Besides, the overall energy consumption reduces very fast after first several iterations. The reason is the proposed SCA-based algorithm is able to converge to a local optimal point after each iteration. Moreover, in this example, the overall energy consumption of the $20$th time slot is much greater than other time slots, while the overall energy consumption of the $10$th and $15$th time slot is lower than other time slot. That is, more energy is required at the first and the last few time slots.

\begin{figure*}[!t]
	\centering 
	\subfloat[]{\includegraphics[width=.5\textwidth]{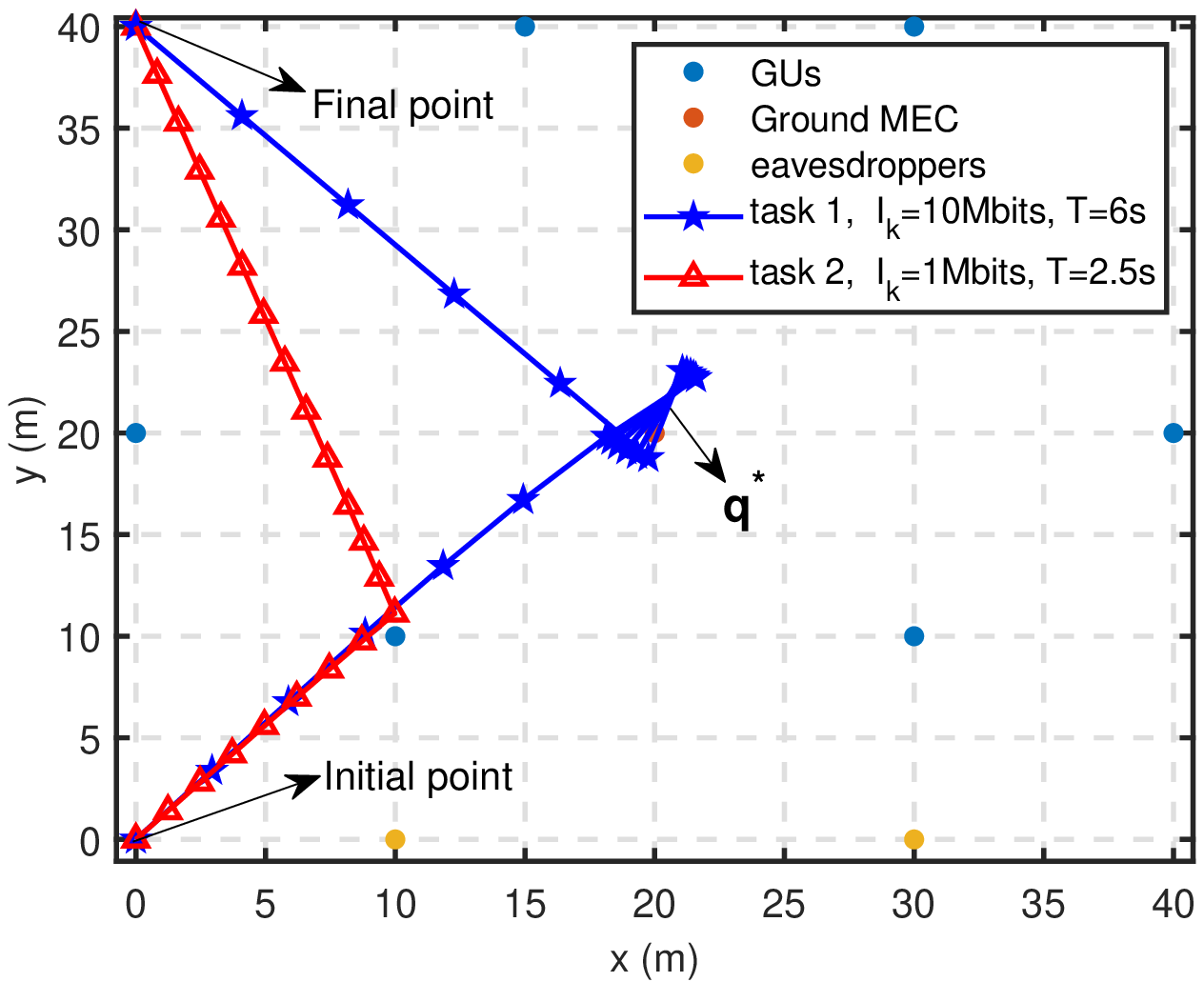}%
		\label{fig:uav_tra}}
	\hfil
	\subfloat[]{\includegraphics[width=.5\textwidth]{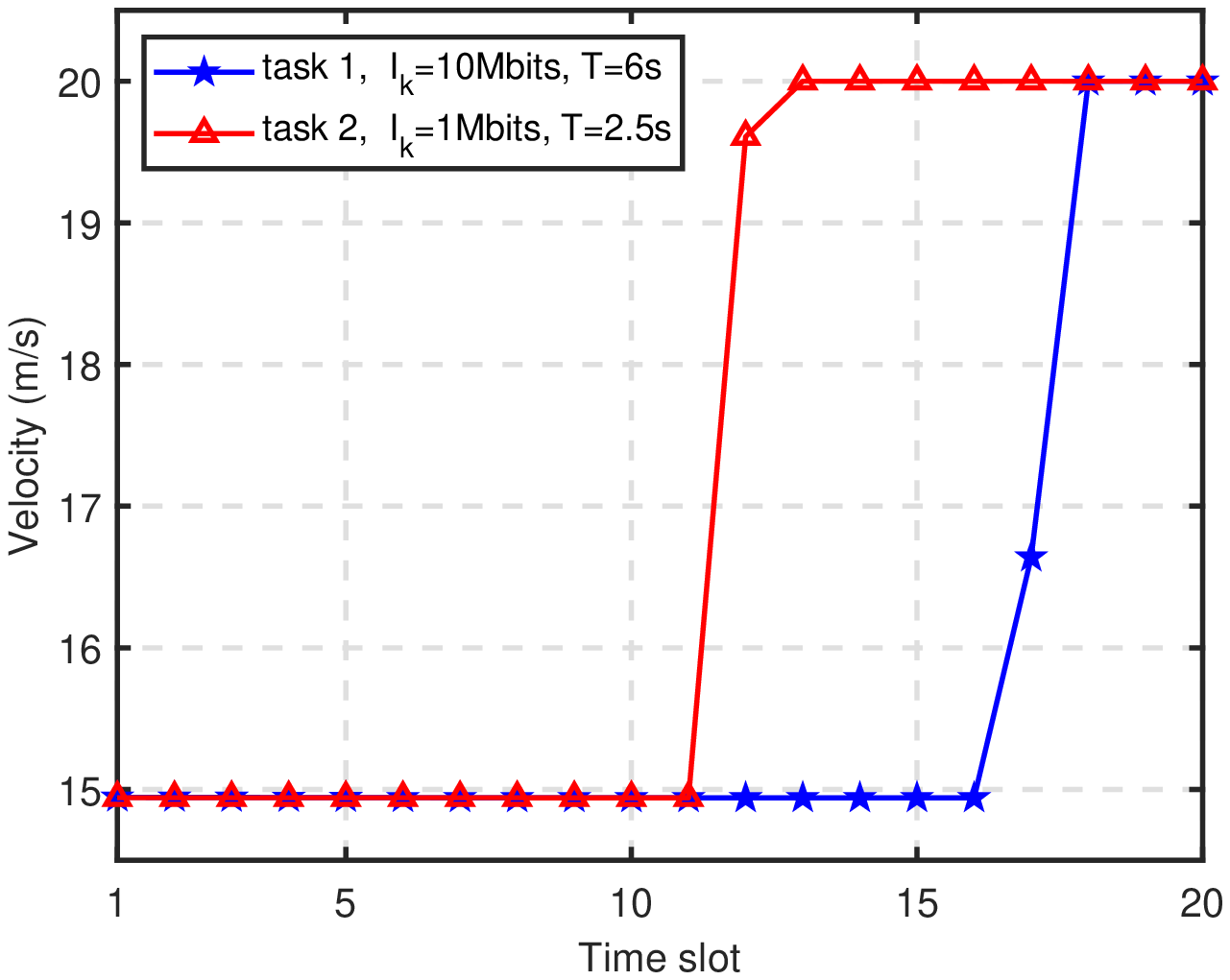}%
		\label{fig:uav_speed}}
	\caption{The UAV trajectory and velocity under two computation tasks: (a) The UAV trajectory, (b) The UAV velocity.}
	\label{fig:uav_tra_speed}
\end{figure*}
\indent Fig. \ref{fig:uav_tra_speed} shows the trajectory and velocity of the UAV under two computation tasks. The computation task $1$ is set as $I_k = 10{\rm Mbits}$ and $T=6{\rm s}$ while the computation task $2$ is set as $I_k = 1{\rm Mbits}$ and $T = 2.5{\rm s}$. From Fig. \ref{fig:uav_tra_speed}(a), it is observed that there exists a horizontal coordinate of UAV with optimal service for GUs, noted as ${\bf q}^{\star}$. For task $1$ with long flight period, the UAV first flies to and then, hovers around ${\bf q}^{\star}$. At last, the UAV flies to ${\bf q}_{\rm F}$ from ${\bf q}^{\star}$. For task $2$ with short flight period, the UAV tries to approach ${\bf q}^{\star}$, while before arriving ${\bf q}^{\star}$, it flies to ${\bf q}_{\rm F}$. From Fig. \ref{fig:uav_tra_speed}(b), it is observed for both tasks, the UAV intends to select a velocity about $14.9{\rm m/s}$ to minimize $P(\Vert {\bf v}[n] \Vert)$, because $P(\Vert {\bf v}[n] \Vert) \Delta_{\rm T}$ dominates the overall energy consumption at each time slot. Once there is limited flight time remained, the UAV speeds up to the maximum velocity in order to arrive at ${\bf q}_{\rm F}$ within the required $T$, and thus, $P({\Vert {\bf v} \Vert})$ is greatly increased, which is consistent with Fig. \ref{fig:uav_con}.
\begin{figure}[t]
	\begin{center}
		\centerline{\includegraphics[ width=.5\textwidth]{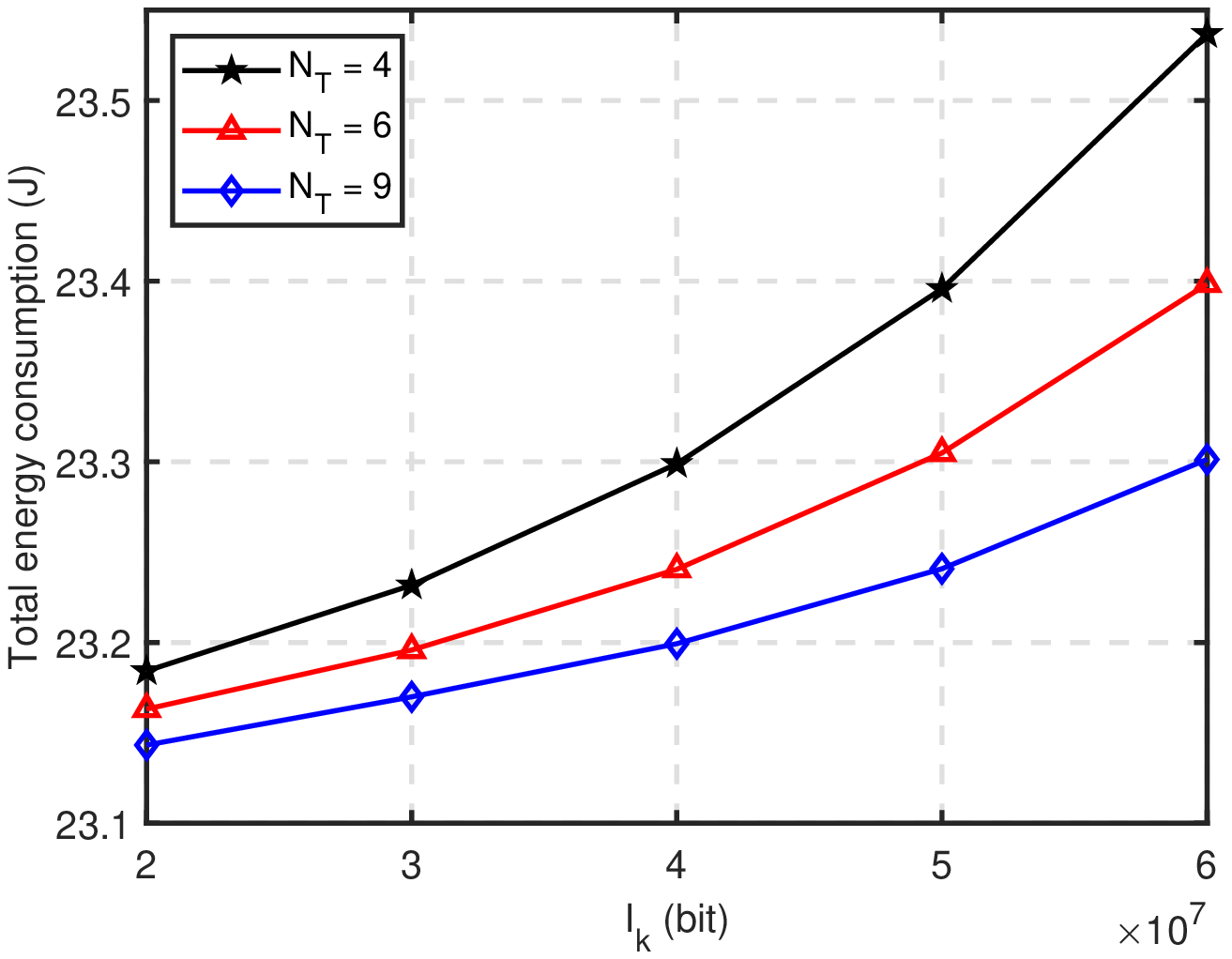}}
		\caption{The total energy consumption versus task input bits $I_k$.}
		\label{fig:uav_ant}
	\end{center}
\end{figure}

\indent 
Denote the sum energy consumption of each time slot as the total energy consumption. Fig. \ref{fig:uav_ant} shows the total energy consumption versus $I_k$, where the results of $N_{\rm T}= 4, 6, 9$ are presented. It is observed that the total energy consumption increases with the increment of $I_k$, and the gap is mainly resulting from energy consumption for task transmission and computation. Besides, with more antennas, the total energy consumption is reduced. Because the increment of antennas provides more spatial degree of freedom to increase the channel gain in both offloading and downloading as well as mitigating the impact of inter-user interference and channel uncertainty. 
\begin{figure}[t]
	\begin{center}
		\centerline{\includegraphics[ width=.5\textwidth]{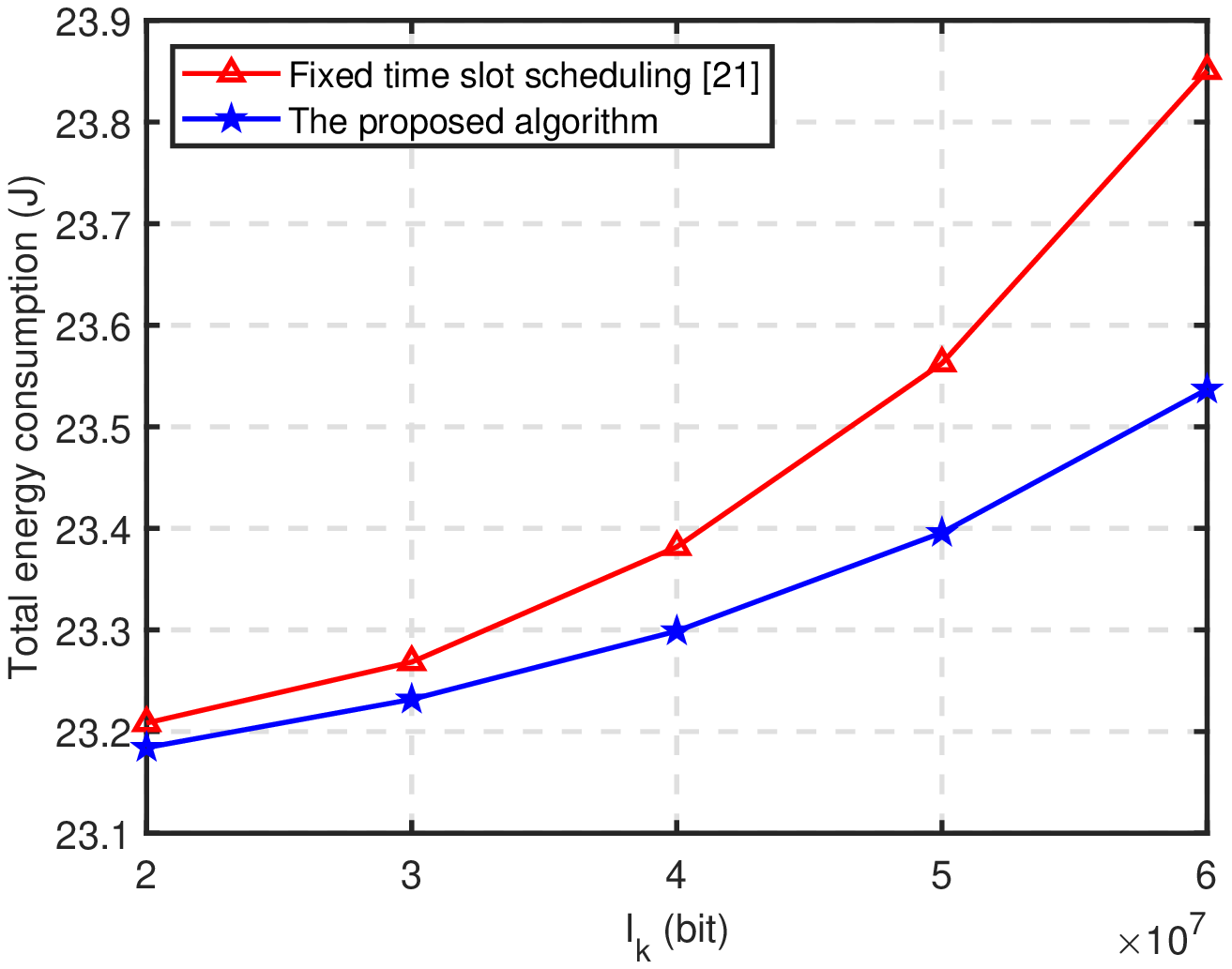}}
		\caption{The total energy consumption versus task input bits $I_k$.}
		\label{fig:uav_tim}
	\end{center}
\end{figure}

\indent Fig. \ref{fig:uav_tim} compares the proposed algorithm with an benchmark scheme adopting fixed time slot scheduling, where for each time slot, the period utilization ratio of the UAV is set as $\theta_{\rm a}=0.5$ while the period  utilization ratios of GUs are identically set as $\theta_k = 0.5/K$ [\ref{xian22}]. It is observed that the proposed algorithm is superior to the benchmark scheme in terms of total energy consumption. The reason is that the optimal solution to the benchmark scheme is in fact a feasible solution to the proposed algorithm. Note that taking time slot scheduling into account, the complex coupling variables make the considered problem more challenging to study. Nevertheless, the proposed algorithm can efficiently handle it.
\begin{figure}[t]
	\begin{center}
		\centerline{\includegraphics[ width=.5\textwidth]{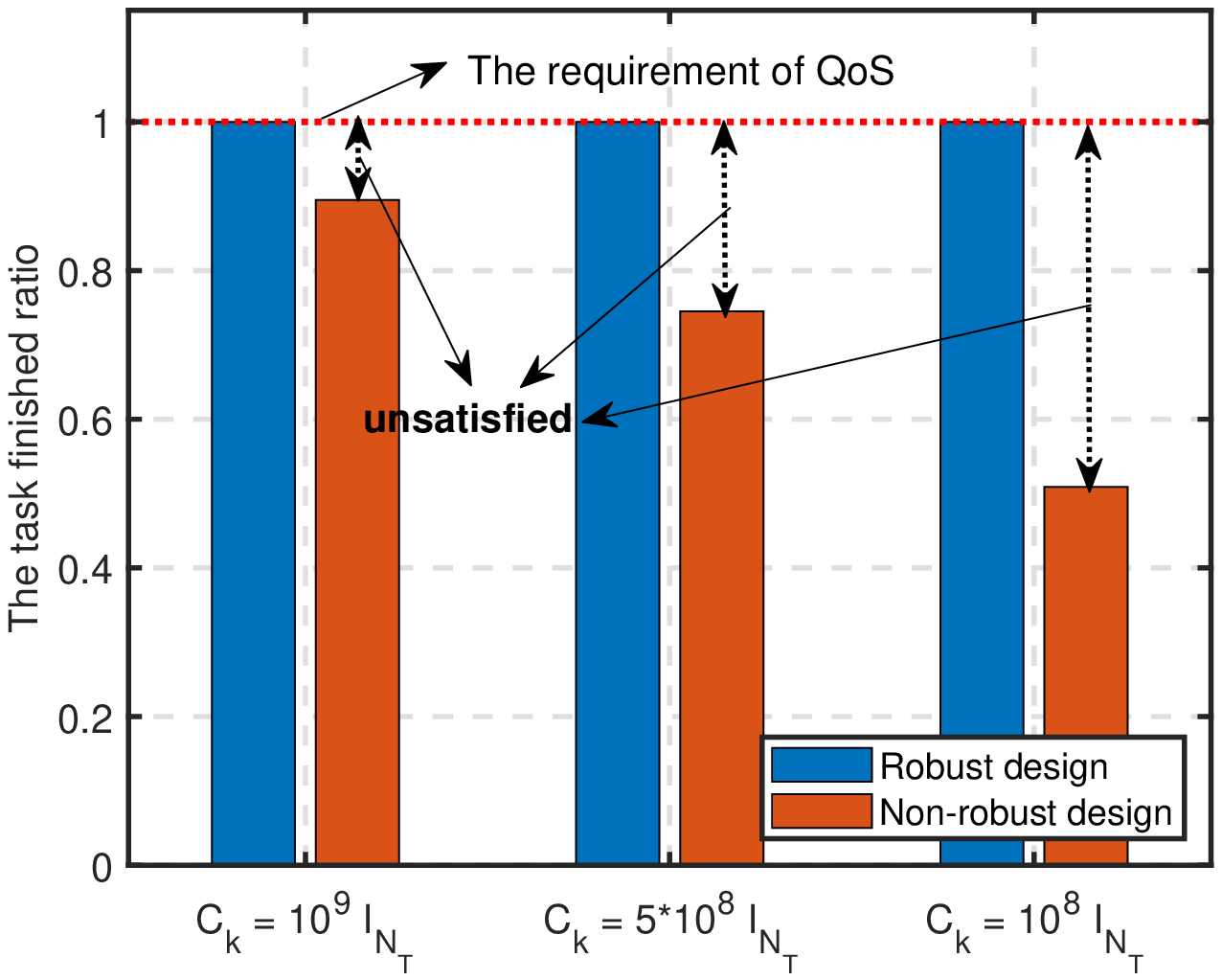}}
		\caption{The task finished ratio in one time slot versus ${\bf C}_k$.}
		\label{fig:uav_rob}
	\end{center}
\end{figure}

\indent 
Fig. \ref{fig:uav_rob} shows the task finished ratio over $10^3$ channel error realizations, where a sample is regarded as a task finished event if the required computation task is finished with given channel error realization. For comparison, the non-robust design is also presented. The results of ${\bf C}_k=10^9{\bf I}_{N_{\rm T}}$, ${\bf C}_k=5*10^8{\bf I}_{N_{\rm T}}$, ${\bf C}_k=10^8{\bf I}_{N_{\rm T}}$ are plotted where the channel error size $\|{\bf e}\|$ is bounded by $1/sqrt(\min {\rm eig}({\bf C}_k))$, and the channel errors are generated based on uniform distribution. It is observed that for considered ${\bf C}_k$, the task finished ratios of the robust design are all $100\%$ which outperforms the non-robust design. Besides, with the decrement of  $\min {\rm eig}({\bf C}_k)$, the task finished ratio of the non-robust design is degraded. In practical wireless communication network, the channel error cannot be neglected. Thus, the robust design is of great significance to guarantee the QoS requirement with channel uncertainty. 

\begin{figure}[t]
	\begin{center}
		\centerline{\includegraphics[ width=.5\textwidth]{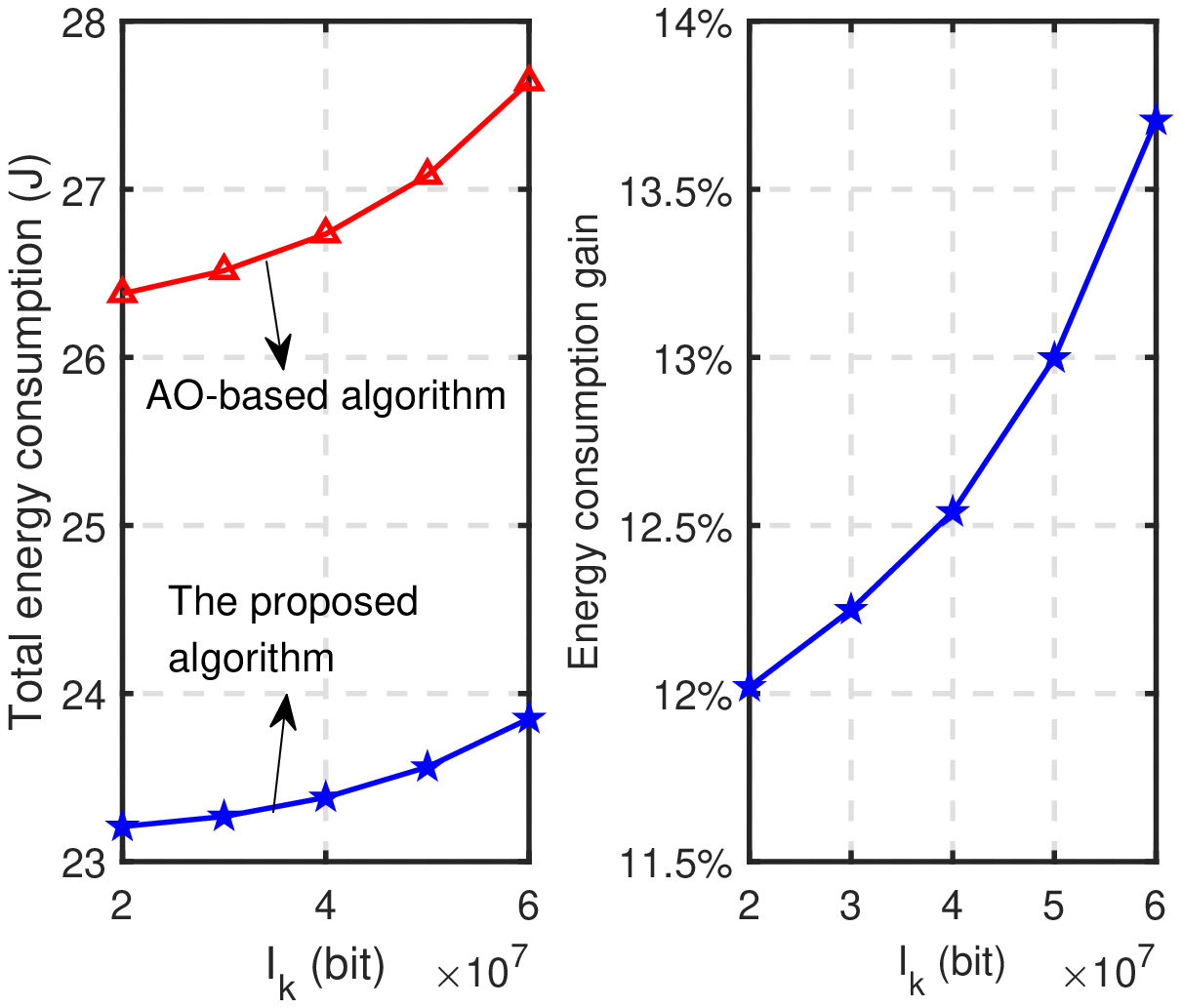}}
		\caption{The total energy consumption and energy consumption gain between the proposed algorithm and AO-based algorithm.}
		\label{fig:uav_ao}
	\end{center}
\end{figure}

\indent
The considered problem can also be solved by traditional AO-based algorithm \cite{mec_relay2, mec_relay3, mec_relay4}. Fig. \ref{fig:uav_ao} compares the proposed algorithm with the AO-based algorithm in terms of the total energy consumption. For comparison, we define a new performance metric named as energy consumption gain, which is computed by $(E_2 - E_1)/E_2$, where $E_1$ and $E_2$ denotes the total energy consumption derived by the proposed algorithm and the AO-based algorithm, respectively. It is observed that the proposed algorithm outperforms the AO-based algorithm, and the energy consumption gain increases with the increment of bits of the computation task ${I}_k$. The reason may be that in each iteration, the AO-based algorithm optimizes the coupling variables separately while the proposed algorithm jointly optimizes all variables after a series of convex reformulations, which consequently yields a superior performance.

\section{Conclusion}
In this paper, energy consumption was minimized for a multi-antenna UAV-assisted MEC network with channel uncertainty in presence of eavesdroppers. An SCA-based algorithm is developed to jointly optimize the computation and transmission resource as well as the trajectory of the UAV, which is shown to be superior to traditional AO-based algorithm numerically. The convergence performance of our developed algorithm was proved and the computation complexity was analyzed. Numerical results validate that our developed scheme outperforms several existing benchmark schemes.


\begin{thebibliography}{1}
	\bibitem{1} \label{xian1}
	R. Zhang, K. Xiong, Y. Lu, B. Gao, P. Fan, and K. B. Letaief, ``Joint coordinated beamforming and power splitting ratio optimization in MU-MISO SWIPT-enabled HetNets: A multi-agent DDQN-based approach,"  {\it IEEE J. Sel. Areas Commun.}, vol. 40, no. 2, pp. 677-693, Feb. 2022.
	
	\bibitem{}\label{xian2}
	J. Zhang et al., ``Computation-efficient offloading and trajectory scheduling for multi-UAV assisted mobile edge computing," {\it IEEE Trans. Veh. Technol.}, vol. 69, no. 2, pp. 2114-2125, Feb. 2020.
	\bibitem{}\label{xian3}
	S. Joo, H. Kang, and J. Kang, ``CoSMoS: Cooperative sky-ground mobile edge computing system," {\it IEEE Trans. Veh. Technol.}, vol. 70, no. 8, pp. 8373-8377, Aug. 2021.
	\bibitem{}\label{xian4}
	Y. Nie, J. Zhao, F. Gao, and F. R. Yu, ``Semi-distributed resource management in UAV-aided MEC systems: A multi-agent federated reinforcement learning approach," {\it IEEE Trans. Veh. Technol.}, vol. 70, no. 12, pp. 13162-13173, Dec. 2021.
	\bibitem{}\label{xian5}
	J. Xu, K. Ota, M. Dong, and H. Zhou, ``MCTS-enhanced hybrid offloading for aerial multi-access edge computing," {\it IEEE Wireless Commun.}, vol. 28, no. 5, pp. 82-87, Oct. 2021.
	\bibitem{1}
	S. Liu et al., ``Satisfaction-maximized secure computation offloading in multi-eavesdropper MEC networks," {\it IEEE Trans. Wireless Commun}, vol. 21, no. 6, pp. 4227-4241, Jun. 2022.		
	\bibitem{2}
	H. Hu, K. Xiong, G. Qu, Q. Ni, P. Fan, and K. B. Letaief, ``AoI-minimal trajectory planning and data collection in UAV-assisted wireless powered IoT networks," {\it IEEE Internet Things J.}, vol. 8, no. 2, pp. 1211-1223, Jan., 2021.
	
	\bibitem{A1}
	 H. Zhang, R. He, B. Ai, S. Cui, and H. Zhang, ``Measuring sparsity of wireless channels”, {\it IEEE Trans. Cogn. Commun. Netw.}, vol. 7, no. 1, pp. 133-144, Mar. 2021.
	 
	
	\bibitem{3} \label{wen9}
	D. Xu, Y. Sun, D. W. K. Ng, and R. Schober, ``Multiuser MISO UAV communications in uncertain environments with no-fly zones: Robust trajectory and resource allocation design," {\it IEEE Trans. Commun}, vol. 68, no. 5, pp. 3153-3172, May 2020.
	\bibitem{4}
	S. Gong, S. Wang, C. Xing, S. Ma, and T. Q. S. Quek, ``Robust superimposed training optimization for UAV assisted communication systems," {\it IEEE Trans. Commun}, vol. 19, no. 3, pp. 1704-1721, Mar. 2020.
	
	\bibitem{A2}
	Z. Ma, B. Ai, R. He, G. Wang, Y. Niu, M. Yang, J. Wang, Y. Li, and Z. Zhong, ``Impact of UAV rotation on MIMO channel characterization for air-to-ground communication systems,” {\it IEEE Trans. Veh. Technol.}, vol. 69, no. 11, pp. 12418-12431, Nov. 2020.
	
	
	\bibitem{}\label{xian14}
	Z. Yang et al., ``AI-driven UAV-NOMA-MEC in next generation wireless networks," {\it IEEE Wireless Commun.}, vol. 28, no. 5, pp. 66-73, Oct. 2021.
	\bibitem{bg1}
	Y. Liu, K. Xiong, Q. Ni, P. Fan, and K. B. Letaief, ``UAV-assisted wireless powered cooperative mobile edge computing: Joint offloading, CPU control, and trajectory optimization," {\it IEEE Internet Things J.}, vol. 7, no. 4, pp. 2777-2790, Apr. 2020.
	
	\bibitem{bg2}
	Y. Yu, X. Bu, K. Yang, H. Yang, X. Gao, and Z. Han, ``UAV-aided low latency multi-access edge computing," {\it IEEE Trans. Veh. Technol.}, vol. 70, no. 5, pp. 4955-4967, May 2021.	
	\bibitem{bg3}
	B. Dai, J. Niu, T. Ren, Z. Hu, and M. Atiquzzaman, ``Towards energy-efficient scheduling of UAV and base station hybrid enabled mobile edge computing," {\it IEEE Trans. Veh. Technol.}, vol. 71, no. 1, pp. 915-930, Jan. 2022.
	\bibitem{mec1}
    H. Peng and X. Shen, ``Multi-agent reinforcement learning based resource management in MEC- and UAV-assisted vehicular networks,"  {\it IEEE J. Sel. Areas Commun.}, vol. 39, no. 1, pp. 131-141, Jan. 2021.
	\bibitem{mec2}
	W. Feng et al., ``Hybrid beamforming design and resource allocation for UAV-aided wireless-powered mobile edge computing networks with NOMA," {\it IEEE J. Sel. Areas Commun.}, vol. 39, no. 11, pp. 3271-3286, Nov. 2021.
	\bibitem{relay1}
	X. Diao, W. Yang, L. Yang, and Y. Cai, ``UAV-relaying-assisted multi-access edge computing with multi-antenna base station: Offloading and scheduling optimization," {\it IEEE Trans. Veh. Technol.}, vol. 70, no. 9, pp. 9495-9509, Sep. 2021.			
	\bibitem{relay2}
	R. Han, Y. Wen, L. Bai, J. Liu, and J. Choi, ``Rate splitting on mobile edge computing for UAV-aided IoT systems," {\it IEEE Trans. Cogn. Commun. Netw.}, vol. 6, no. 4, pp. 1193-1203, Dec. 2020.
	
	
	
	\bibitem{mec_relay1}
	Z. Yu, Y. Gong, S. Gong, and Y. Guo, ``Joint task offloading and resource allocation in UAV-enabled mobile edge computing,"  {\it IEEE Internet Things J.}, vol. 7, no. 4, pp. 3147-3159, Apr. 2020.
	
	\bibitem{mec_relay2} \label{wen6}
	X. Hu, K.Wong, K. Yang, and Z. Zheng, ``UAV-assisted relaying and edge computing: Scheduling and trajectory optimization,” {\it IEEE Trans.Wireless Commun.}, vol. 18, no. 10, pp. 4738–4752, Oct. 2019.
	\bibitem{mec_relay3}
	X. Hu, K. -K. Wong, and Y. Zhang, ``Wireless-powered edge computing with cooperative UAV: Task, time scheduling and trajectory design," {\it IEEE Trans.Wireless Commun.}, vol. 19, no. 12, pp. 8083-8098, Dec. 2020.
	\bibitem{mec_relay4} \label{xian22}
	B. Liu, Y. Wan, F. Zhou, Q. Wu, and R. Q. Hu, ``Resource allocation and trajectory design for MISO UAV-assisted MEC networks," {\it IEEE Trans. Veh. Technol.}, vol. 71, no. 5, pp. 4933-4948, May 2022.
	
	\bibitem{sec1}
	Y. Xu, T. Zhang, D. Yang, Y. Liu, and M. Tao, ``Joint resource and trajectory optimization for security in UAV-assisted MEC systems," {\it IEEE Trans. Commun.}, vol. 69, no. 1, pp. 573-588, Jan. 2021.
	\bibitem{sec2} 
	W. Lu et al., ``Secure NOMA-based UAV-MEC network towards a flying eavesdropper," {\it IEEE Trans. Commun.}, vol. 70, no. 5, pp. 3364-3376. May 2022.
	\bibitem{sec3} 
	Y. Zhou et al., ``Secure communications for UAV-enabled mobile edge computing systems," {\it IEEE Trans. Commun.}, vol. 68, no. 1, pp. 376-388, Jan. 2020.
	\bibitem{sec4}
	Y. Yapıcı, N. Rupasinghe, İ. Güvenç, H. Dai, and A. Bhuyan, ``Physical layer security for NOMA transmission in mmWave drone networks," {\it IEEE Trans. Veh. Technol.}, vol. 70, no. 4, pp. 3568-3582, Apr. 2021.
	
	
	
	
		
	\bibitem{rob1}
	H. Zhang, J. Zhang, and K. Long, ``Energy efficiency optimization for NOMA UAV network with imperfect CSI," {\it IEEE J. Sel. Areas Commun.}, vol. 38, no. 12, pp. 2798-2809, Dec. 2020.
	\bibitem{rob3}
	Y. Li, H. Zhang, and K. Long, ``Joint resource, trajectory, and artificial noise optimization in secure driven 3-D UAVs with NOMA and imperfect CSI," {\it IEEE J. Sel. Areas Commun.}, vol. 39, no. 11, pp. 3363-3377, Nov. 2021.
	\bibitem{rob4}
	Y. Lu, Y. Huang, and T. Hu, ``Robust resource scheduling for air-ground cooperative mobile edge computing," {\it IEEE ICCC}, pp. 764-769, 2021.
	




	
	
	

	
	
	

	


	\bibitem{}\label{wen8}
	Y. Zeng, J. Xu, and R. Zhang, ``Energy minimization for wireless communication with rotary-Wing UAV,” {\it IEEE Trans. Wireless Commun.}, vol. 18, no. 4, pp. 2329-2345, Apr. 2019.
	\bibitem{xian10}
	S. Li, B. Duo, M. D. Renzo, M. Tao.l, and X. Yuan, ``Robust secure UAV communications with the aid of reconfigurable intelligent surfaces," {\it IEEE Trans. Wireless Commun.}, vol. 20, no. 10, pp. 6402-6417, Oct. 2021.
	
	
	
	
	\bibitem{} \label{wen3}
	 Q. Wu, Y. Zeng, and R. Zhang, ``Joint trajectory and communication design for Multi-UAV enabled wireless networks,” {\it IEEE Trans. Wireless Commun.}, vol. 17, no. 3, pp. 2109-2121, Mar. 2018

	 \bibitem{} \label{wen4}
	 Y. Zhou, F. Zhou, H. Zhou, D. W. K. Ng, and R. Q. Hu, ``Robust trajectory and transmit power optimization for secure UAV-enabled cognitive radio networks,” {\it IEEE Trans. Commun.}, vol. 22, no. 1, pp.161-164, Jan. 2018.
	 \bibitem{} \label{wen5}
	 Y. Cai, Z. Wei, R. Li, D. W. K. Ng, and J. Yuan, ``Joint trajectory and resource allocation design for energy-efficient secure UAV communication systems,” {\it IEEE Trans. Commun.}, vol. 68, no. 7, pp. 4536-4553, Jul. 2020.
	  \bibitem{xian11}
	  X. Qin, Z. Song, Y. Hao, and X. Sun, ``Joint resource allocation and trajectory optimization for multi-UAV-assisted multi-access mobile edge computing," {\it IEEE Wireless Commun. Lett. }, vol. 10, no. 7, pp. 1400-1404, Jul. 2021.
	  \bibitem{z1}
	  H. Mei, K. Yang, J. Shen, and Q. Liu, ``Joint trajectory-task-cache optimization with phase-shift design of RIS-assisted UAV for MEC," {\it IEEE Wireless Commun. Lett. }, vol. 10, no. 7, pp. 1586-1590, Jul. 2021.
	  \bibitem{xian12}
	  L. V. S. Boyd, Convex Optimization. Cambridge, U.K.: Cambridge Univ.Press, Mar. 2004.
	  \bibitem{alg1}
	  A. Ben-Tal and A. Nemirovski, Lectures on Modern Convex Optimization: Analysis, Algorithms, and Engineering Applications. Philadelphia, PA, USA: SIAM, 2001.
	  \bibitem{alg2}
	  G. Zhou, C. Pan, H. Ren, K. Wang, and A. Nallanathan, ``A framework of robust transmission design for IRS-aided MISO communications with imperfect cascaded channels,” {\it IEEE Trans. Signal Process.}, vol. 68, pp. 5092–5106, 2020.
\end{thebibliography}
\end{document}